\documentclass[lettersize,journal]{IEEEtran}
\usepackage{amsmath,amsfonts}
\usepackage{array}
\usepackage{textcomp}
\usepackage{stfloats}
\usepackage{url}
\usepackage{tikz}
\usetikzlibrary{arrows.meta,positioning,fit,calc}
\usepackage{adjustbox}
\usepackage{verbatim}
\usepackage{comment}
\usepackage{cite}
\usepackage{bm}  
\usepackage{float} 
\usepackage{placeins} 
\usepackage{amsmath,amssymb,amsfonts}
\usepackage{graphicx}
\usepackage{textcomp}
\usepackage{xcolor}
\usepackage{caption}
\usepackage{cite}
\usepackage{multirow}
\usepackage{url}
\usepackage{amsmath,amssymb,amsfonts}
\usepackage{graphicx}
\usepackage{textcomp}
\usepackage{caption}
\usepackage{multicol}
\usepackage{xcolor}
\usepackage{subcaption}
\usepackage{array}
\usepackage{float}
\usepackage{hyperref}
\usepackage{algorithm}
\usepackage{algorithmic}
\usepackage{graphicx}
\usepackage{textcomp}
\usepackage{scalerel}
\usepackage{cite}
\usepackage{amsmath,amssymb,amsfonts}
\usepackage{algorithmic}
\usepackage{graphicx}
\usepackage{textcomp}
\usepackage{caption}
\usepackage{multicol}
\usepackage{xcolor}
\usepackage{esdiff}
\usepackage{cleveref}
\usepackage{etoolbox}
\usepackage{cases}
\usepackage{mathtools}
\usepackage{comment}

\newtheorem{proof}{Proof}
\renewcommand{\baselinestretch}{1.025}
\usepackage{soul}

\usepackage[none]{hyphenat} 
\usepackage{microtype}      
\begin{document}
	
	\title{D²-UC: A Distributed–Distributed Quantum–Classical Framework for Unit Commitment}
	
	\author{Milad Hasanzadeh  \textit{Graduate Student Member, IEEE} and Amin Kargarian, \textit{Senior Member, IEEE}
		\thanks{This work was supported by the National Science Foundation under Grant ECCS-1944752 and Grant ECCS-2312086.
			
			The authors are with the Electrical and Computer Engineering Department, Louisiana State University, Baton Rouge, LA 70803 USA (email: mhasa42@lsu.edu, kargarian@lsu.edu).}}

	\maketitle
 
\begin{abstract}
This paper introduces D²-UC, a quantum-ready framework for the unit commitment (UC) problem that prepares UC for near-term hybrid quantum–classical solvers by combining distributed classical decomposition with distributed quantum execution. We reformulate deterministic and stochastic UC into a three-block alternating direction method of multipliers (ADMM): (i) a convex quadratic subproblem for dispatch and reserves, (ii) a binary subproblem expressed as a quadratic unconstrained binary optimization (QUBO), and (iii) a proximal slack update for consensus. The core contributions are fivefold. First, we demonstrate how the full UC problem can be expressed as a single, monolithic QUBO, thereby establishing a direct interface to quantum solvers. Second, we decompose this large binary block into three type-specific QUBOs for commitment, startup, and shutdown, making the problem more tractable but revealing slower ADMM convergence. Third, we restore local logical couplings through per unit–time micro-QUBOs, which accelerate convergence. Fourth, we batch micro-QUBOs into $K$ non-overlapping block-diagonal problems, reducing a large number of subproblems to a fixed number of solver-ready QUBOs per iteration, well matched to distributed variational quantum eigensolvers (DVQE). Fifth, we integrate an accept-if-better safeguard with DVQE to stabilize hybrid updates and prevent oscillations. Case studies on deterministic and stochastic UC confirm that the proposed methods deliver feasible schedules, faster convergence, and QUBO sizes aligned with current and near-term quantum hardware capabilities. \ul{All detailed data, codes, and parameter values for all cases studied are available at} \href{https://github.com/LSU-RAISE-LAB/3B-ADMM-UC-DVQE}{GitHub}.  
\end{abstract}


\section*{Nomenclature}
\addcontentsline{toc}{section}{Nomenclature}
\begin{IEEEdescription}[
  \IEEEusemathlabelsep\IEEEsetlabelwidth{$r^{\uparrow}_{i,t,s}, r^{\downarrow}_{i,t,s}$}]

\item[\emph{Indices, Sets, and Functions:}]
\item[$i$] Index of generating units, $i \in \{1,\dots,N\}$.
\item[$t$] Index of time periods, $t \in \{1,\dots,T\}$.
\item[$s$] Index of uncertainty scenarios, $s \in \{1,\dots,S\}$.
\item[$k$] Index for ADMM iterations.

\item[\emph{Parameters:}]
\item[$L_t$] System load demand at time $t$ [MW].
\item[$\tilde{L}_{t,s}$] Net load at time $t$ under scenario $s$ [MW].
\item[$P_i^{\min}, P_i^{\max}$] Minimum/maximum generation capacity of unit $i$ [MW].
\item[$R_t^{\uparrow}, R_t^{\downarrow}$] Up/downward reserve requirements at time $t$ [MW].
\item[$R_{t,s}^{\uparrow}, R_{t,s}^{\downarrow}$] Scenario-dependent reserve requirements at time $t$ [MW].
\item[$RU_i, RD_i$] Ramp-up/-down limits of unit $i$ [MW/h].
\item[$SU_i, SD_i$] Startup and shutdown ramp limits [MW/h].
\item[$U_i, D_i$] Minimum up and down time requirements [h].
\item[$A_i, B_i, C_i$] Fixed, linear, and quadratic cost coefficients.
\item[$S_i, H_i$] Startup and shutdown cost coefficients.
\item[$\Delta\tau$] Reserve response time window [h].
\item[$\pi_s$] Probability of scenario $s$.
\item[$\rho_y, \rho_u, \rho_v$] ADMM penalty parameters for $y,u,v$.
\item[$\beta_y, \beta_u, \beta_v$] Proximal penalty weights for $y,u,v$.
\item[$\gamma_c, \gamma_{ss},$] Micro-QUBO penalty weights.  
\item[$\gamma_{u\to y}, \gamma_{v\to\bar y}$] Micro-QUBO penalty weights.
\item[$\gamma_y, \gamma_u, \gamma_v$] Anchoring penalty weights for relaxed Block~1.

\item[\emph{Variables:}]
\item[$y_{i,t}$] Commitment status of unit $i$ at time $t$.
\item[$u_{i,t}$] Startup indicator for unit $i$ at time $t$.
\item[$v_{i,t}$] Shutdown indicator for unit $i$ at time $t$.
\item[$p_{i,t,s}$] Generation output of unit $i$ at time $t$ under scenario $s$ [MW].
\item[$r^{\uparrow}_{i,t}, r^{\downarrow}_{i,t}$] Up/downward reserves of unit $i$ at time $t$ [MW].
\item[$r^{\uparrow}_{i,t,s}, r^{\downarrow}_{i,t,s}$] Up/downward reserves under scenario $s$ [MW].
\item[$z^y_{i,t}, z^u_{i,t}, z^v_{i,t}$] Auxiliary binary proxies for $y,u,v$.
\item[$\xi^y_{i,t}, \xi^u_{i,t}, \xi^v_{i,t}$] Proximal slack variables for $y,u,v$.
\item[$\lambda^y_{i,t}, \lambda^u_{i,t}, \lambda^v_{i,t}$] Dual variables associated with consensus constraints.
\item[$z_{i,t}$] Binary vector $[z^y_{i,t}, z^u_{i,t}, z^v_{i,t}]^\top \in \{0,1\}^3$.
\item[$E_{i,t}(z)$] Local micro-QUBO energy function at $(i,t)$.
\end{IEEEdescription}

\section{Introduction}\label{sec:introduction}

\IEEEPARstart{U}{nit} commitment (UC) is a core optimization problem in power system operations, determining unit on/off status to minimize cost while satisfying demand, reserve, and technical constraints \cite{padhy2004unit}. Formulated as a mixed integer linear programming/mixed integer quadratic programming (MILP/MIQP) and proven NP-hard \cite{carrion2006computationally}, UC remains challenging despite advances in solvers and decomposition. Growing system scale and renewable integration further amplify its dimensionality, uncertainty, and computational burden \cite{asensio2015stochastic, mahroo2023learning, nikmehr2022quantum,dong2025data}. The complexity of UC grows rapidly with system size, temporal detail, and uncertainty, making large-scale MILP/MIQP hard for classical solvers \cite{javadi2025learning}. Quantum computing offers a promising alternative for such combinatorial tasks \cite{farhi2014quantum, preskill2018quantum,pareek2025limitations,morstyn2022annealing}, with hybrid quantum--classical algorithms like variational quantum eigensolver (VQE) \cite{peruzzo2014variational} and quantum approximate optimization algorithm (QAOA) \cite{farhi2014quantum} emerging as practical near-term approaches for noisy intermediate-scale quantum (NISQ) devices.

Hybrid quantum--classical algorithms target quadratic unconstrained binary optimization (QUBO) problems, while UC is a mixed-integer program with both binary and continuous variables. Decomposition methods such as Benders and alternating direction method of multipliers (ADMM) \cite{conejo2006decomposition, zhang2020two,11177244,shi2025two} separate UC into convex continuous subproblems (dispatch, reserves) solvable classically and a discrete subproblem of binary commitments, which can be reformulated as a QUBO for quantum solvers \cite{han2025quantum,mahroo2023learning}. A three-block ADMM decomposition aligns UC with quantum solvers by separating continuous and discrete decisions \cite{gambella2020multiblock}. Block~1 solves a convex QP for dispatch and reserves. Block~2 reformulates binary commitments, startups, and shutdowns as a QUBO with auxiliary proxies and slack variables. Block~3 updates the slack variables to enforce consensus. This structure preserves UC’s physical and operational constraints in a tractable form \cite{zhang2020two,stein2305combining}.

Despite their promise, current quantum devices remain in the NISQ era, with limited numbers of qubits, restricted connectivity, and short coherence times \cite{preskill2018quantum}. As a result, solving large QUBO instances—such as those arising from full-scale UC formulations—remains a challenge, even when using hybrid quantum–classical algorithms. The gap between the problem sizes of practical interest and the hardware capabilities of existing quantum processors makes direct application infeasible \cite{nikmehr2022quantum,mahroo2023learning}. Two promising avenues to mitigate these limitations are QUBO decomposition methods, which partition a large QUBO into multiple smaller, tractable subproblems, and distributed quantum computing, which leverages parallel quantum resources or circuit partitioning to jointly address larger optimization tasks \cite{glover2018tutorial, caleffi2024distributed, hasanzadeh2025distributed}. 

While the UC problem is formulated as a system-wide optimization, its binary structure exhibits a high degree of separability. In particular, the commitment, startup, and shutdown variables of each generating unit evolve independently of those of other units; for example, whether one unit starts up or remains committed has no direct effect on the binary status of another unit. The interdependence among units arises primarily through continuous constraints such as power balance, ramping limits, and reserve requirements, which can be handled in the convex subproblems \cite{carrion2006computationally,padhy2004unit}. This structural property has also been recognized in emerging quantum-inspired formulations of UC \cite{nikmehr2022quantum}, and it enables the decomposition of a large monolithic QUBO—containing all binary decisions across all units and periods—into multiple smaller QUBOs. Such a decomposition aligns naturally with the capabilities of current hybrid quantum–classical algorithms, allowing them to process these smaller QUBOs.

A further avenue to overcome the qubit limitations of current hardware is distributed quantum computing, where multiple small quantum processors cooperate to solve a larger optimization task. Recent advances, such as the TeleGate protocol, enable the execution of quantum gates across spatially separated quantum devices, creating the illusion of a single larger QPU by using entanglement and classical communication \cite{du2022distributed,diadamo2021distributed,yimsiriwattana2004distributed}. Building on such distributed primitives, hybrid algorithms can be extended to distributed settings. For example, the distributed variational quantum eigensolver (DVQE) has been proposed as a distributed version of VQE, designed to solve QUBO problems by partitioning them across multiple quantum processors and aggregating results \cite{hasanzadeh2025distributed}. In DVQE, each QPU executes a subset of parameterized quantum circuits, and classical coordination iteratively updates the shared variational parameters. This architecture allows near-term QPUs to jointly tackle QUBOs of a size that would exceed the capacity of any single processor, making distributed quantum computing particularly promising for large-scale decompositions of the UC problem.

\begin{figure}[!t]
\centering

\definecolor{LeftCol}{RGB}{42,120,217}   
\definecolor{MidCol}{RGB}{0,128,109}     
\definecolor{RightCol}{RGB}{226,138,0}   
\definecolor{DashCol}{RGB}{110,110,110}
\definecolor{Ink}{RGB}{25,25,25}

\resizebox{\columnwidth}{!}{%
\begin{tikzpicture}[
  font=\footnotesize\sffamily,
  >={Latex[length=1.3mm]},
  line width=0.45pt,
  block/.style={draw=Ink, rounded corners=1.6pt, align=center,
                minimum width=26mm, minimum height=6mm,
                inner ysep=2pt, inner xsep=2.6pt, text depth=0.4ex, font=\scriptsize},
  Lblock/.style={block, fill=LeftCol!8,  draw=LeftCol!60!black},
  Mblock/.style={block, fill=MidCol!8,   draw=MidCol!60!black},
  Rblock/.style={block, fill=RightCol!8, draw=RightCol!60!black, minimum width=24mm},
  group/.style={draw=DashCol, dashed, rounded corners=4pt, inner sep=2pt, line width=0.45pt},
  note/.style={font=\scriptsize, text=DashCol},
  link/.style={dashed, draw=DashCol, line width=0.45pt}
]

\def\colsep{7mm}

\node[Lblock] (dispatch) {Dispatch \& Reserves\\(Convex QP)};
\node[Mblock, below=2mm of dispatch] (qubo) {QUBO Problem};
\node[Lblock, below=2mm of qubo] (prox) {Proximal Slacks\\(Unconstrained)};
\node[Lblock, below=2mm of prox] (dual) {Dual Variable Updates};

\draw[purple!70!black,->] (dispatch) -- (qubo);
\draw[purple!70!black,->] (qubo) -- (prox);
\draw[purple!70!black,->] (prox) -- (dual);

\draw[purple!70!black,->]
  ([yshift=-0.5mm]dual.south) to[bend right=120]
  ([yshift=0.5mm]dispatch.north);

\node[group,fit=(dispatch)(dual)] (grpL) {};
\node[note, font=\tiny] at ([yshift=-1.2mm]grpL.south) {ADMM outer loop};

\node[Rblock, right=\colsep of dispatch] (dec) {QUBO Decompositions};
\node[Mblock, below=2mm of dec] (vqe) {Parallel VQE solvers\\on multiple QPUs};
\node[Mblock, below=2mm of vqe] (adam) {ADAM solver\\for ansatz updates};
\node[Mblock, below=2mm of adam] (acc) {Accept if better\\safeguard};

\draw[purple!70!black,->] (dec) -- (vqe);

\draw[purple!70!black,->] ($(vqe.south)+(1.0mm,0)$) -- ($(adam.north)+(1.0mm,0)$);
\draw[purple!70!black,->] ($(adam.north)-(1.0mm,0)$) -- ($(vqe.south)-(1.0mm,0)$);

\draw[purple!70!black,->] (adam) -- (acc);

\node[group,fit=(dec)(acc)] (grpM) {};
\node[note, font=\tiny] at ([yshift=-1.2mm]grpM.south) {DVQE inner loop (QUBO Problem)};

\node[text=MidCol!80!black] at ($($(qubo.east)!0.55!(grpM.west)$)+(0,2mm)$) {\Large $\Rightarrow$};

\node[Rblock, right=\colsep of dec] (t) {Three QUBOs};
\node[Rblock, below=1.8mm of t] (m) {Micro QUBOs};
\node[Rblock, below=1.8mm of m] (b) {Batched QUBOs};

\node[group,fit=(t)(b)] (grpR) {};
\node[note, font=\tiny] at ([yshift=-1.2mm]grpR.south) {3 Decompositions};

\node[text=RightCol!80!black] at ($($(dec.east)!0.55!(grpR.west)$)+(0,4mm)$) {\Large $\Rightarrow$};

\end{tikzpicture}%
}

\caption{Overview of the proposed framework}
\label{overviewFig}
\end{figure}
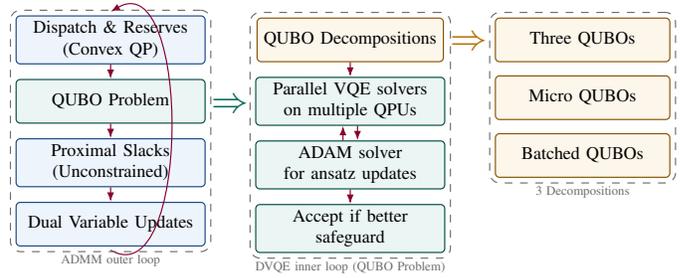

To address the challenges discussed above, we introduce D²-UC, a quantum-ready framework for UC that prepares the problem for near-term hybrid quantum–classical solvers by combining distributed classical decomposition with distributed quantum execution. We present three complementary decomposition strategies for isolating and solving the binary component of UC within a quantum-ready framework. The first approach decomposes the large QUBO into three independent blocks, separating the binary variables associated with unit commitments, startups, and shutdowns. The second approach restructures the problem into micro-QUBOs, each defined per unit–time pair, which restores local couplings and improves ADMM convergence. Building on this idea, a third strategy groups micro-QUBOs into batched QUBOs based on a hardness measure, a tool commonly used to characterize rugged energy landscapes in quantum optimization problems. To solve QUBOs of appropriate size, the framework leverages the DVQE, a distributed extension of VQE that enables QUBO problems to be partitioned and executed across multiple quantum processors. In addition, a safeguard mechanism is proposed to mitigate the “ping-pong” behavior reported in prior studies of hybrid quantum–classical methods, including observations made when applying DVQE. Fig. \ref{overviewFig} illustrates the D$^{2}$-UC framework, comprising the ADMM outer loop and the DVQE inner loop.

The paper is organized as follows. Section~\ref{sec:uc_admm} formulates stochastic unit commitment and introduces the three-block ADMM decomposition. Section~\ref{sec:Three_block_ADMM_Con} establishes Lyapunov-based stability for the three-block scheme. Section~\ref{sec:qubo_dvqe} develops the QUBO decompositions (three, micro, batched) and details the DVQE integration. Section~\ref{sec:results} reports deterministic and stochastic case studies, and Section~\ref{sec:conclusion} provides concluding remarks. \textbf{\ul{All detailed data, codes, and parameter values for all cases studied are available at} \href{https://github.com/LSU-RAISE-LAB/3B-ADMM-UC-DVQE}{GitHub}. }

\section{UC Decomposition with Three-Block ADMM}\label{sec:uc_admm}

We formulate stochastic UC with renewable uncertainty and present a three-block ADMM decomposition that partitions the problem into continuous dispatch/reserves, binary QUBO, and proximal slack/dual updates.

\subsection{Stochastic UC Formulation}\label{sec:sto_uc}

Uncertainty in UC arises from variability in renewable generation and load forecasts. We adopt a scenario-based stochastic framework in which the net load at each time period is represented by a finite set of realizations with probabilities $\pi_s$. First-stage binary variables $(y,u,v)$ are nonanticipative across scenarios, while second-stage continuous variables $(p,r^{\uparrow},r^{\downarrow})$ adapt to each scenario. The stochastic UC with $N$ units is formulated as (1). The deterministic UC is recovered when $S=1$.
\begin{subequations}\label{eq:sto_constraints}
\begin{align}
\min \ &
\sum_{t=1}^{T}\sum_{i=1}^{N}\!\Big(A_i y_{i,t} + S_i u_{i,t} + H_i v_{i,t}\Big) \nonumber\\
&+ \sum_{s=1}^{S} \pi_s \sum_{t=1}^{T}\sum_{i=1}^{N}\!\Big(B_i p_{i,t,s} + C_i p_{i,t,s}^2\Big)
\tag{\ref{eq:sto_constraints}a}\label{eq:sto_obj} \\[4pt]
\text{s.t. } \nonumber\\
& \sum_{i=1}^N p_{i,t,s} = \tilde{L}_{t,s}, \qquad \qquad \qquad \qquad \forall t,s \tag{\ref{eq:sto_constraints}b} \\
& P_i^{\min} y_{i,t} \le p_{i,t,s} \le P_i^{\max} y_{i,t}, \qquad \quad \forall i,t,s \tag{\ref{eq:sto_constraints}c} \\
& y_{i,t} - y_{i,t-1} = u_{i,t} - v_{i,t}, \qquad \qquad ~~\forall i,t \tag{\ref{eq:sto_constraints}d} \\
& u_{i,t} + v_{i,t} \le 1, \qquad \qquad \qquad \qquad \quad ~\forall i,t \tag{\ref{eq:sto_constraints}e} \\
& \sum_{k=\max\{1,t-U_i+1\}}^t u_{i,k} \le y_{i,t}, \qquad \qquad \forall i,t \tag{\ref{eq:sto_constraints}f} \\
& \sum_{k=\max\{1,t-D_i+1\}}^t v_{i,k} \le 1-y_{i,t}, \qquad \forall i,t \tag{\ref{eq:sto_constraints}g} \\
& p_{i,t,s} - p_{i,t-1,s} \le RU_i y_{i,t-1} + SU_i u_{i,t},  \quad \forall i,t,s \tag{\ref{eq:sto_constraints}h} \\
& p_{i,t-1,s} - p_{i,t,s} \le RD_i y_{i,t} + SD_i v_{i,t}, \quad ~~\forall i,t,s \tag{\ref{eq:sto_constraints}i} \\
& 0 \le r_{i,t,s}^{\uparrow} \le P_i^{\max} y_{i,t} - p_{i,t,s}, \qquad \qquad \forall i,t,s \tag{\ref{eq:sto_constraints}j} \\
& 0 \le r_{i,t,s}^{\downarrow} \le p_{i,t,s} - P_i^{\min} y_{i,t}, \qquad \qquad ~\forall i,t,s \tag{\ref{eq:sto_constraints}k} \\
& r_{i,t,s}^{\uparrow} \le RU_i \Delta\tau,\quad 
r_{i,t,s}^{\downarrow} \le RD_i \Delta\tau, \qquad \forall i,t,s \tag{\ref{eq:sto_constraints}l} \\
& \sum_{i=1}^N r_{i,t,s}^{\uparrow} \ge R_{t,s}^{\uparrow}, \quad
\sum_{i=1}^N r_{i,t,s}^{\downarrow} \ge R_{t,s}^{\downarrow}, \qquad \forall t,s \tag{\ref{eq:sto_constraints}m} \\
& y_{i,t},u_{i,t},v_{i,t}\in\{0,1\}, \;
p_{i,t,s},r_{i,t,s}^{\uparrow},r_{i,t,s}^{\downarrow}\ge 0, \forall i,t,s \tag{\ref{eq:sto_constraints}n}
\end{align}
\end{subequations}

Power balance matches generation to net load as in (\ref{eq:sto_constraints}b). Bounds, logic, minimum up/down times, and ramping are enforced in (\ref{eq:sto_constraints}c)–(\ref{eq:sto_constraints}i). Reserve deliverability and adequacy are enforced by (\ref{eq:sto_constraints}j)–(\ref{eq:sto_constraints}m). Domain constraints are given in (\ref{eq:sto_constraints}n).

\subsection{Three-Block ADMM Decomposition}\label{sec:3block_admm}

To interface UC with binary QUBO solvers, we decouple integrality by introducing auxiliary binary proxies $Z=\{z^y,z^u,z^v\}$ and one-sided proximal slacks $\Xi=\{\xi^y,\xi^u,\xi^v\}$, replacing (\ref{eq:sto_constraints}n) with consensus relations \cite{gambella2020multiblock}.

\subsubsection*{Consensus Constraints}

For all $i,t$,
\begin{subequations}\label{eq:consensus_constraints}
\begin{align}
& 0 \le y_{i,t},\,u_{i,t},\,v_{i,t} \le 1, \\
& y_{i,t} - z^y_{i,t} + \xi^y_{i,t} = 0, \\
& u_{i,t} - z^u_{i,t} + \xi^u_{i,t} = 0, \\
& v_{i,t} - z^v_{i,t} + \xi^v_{i,t} = 0, \\
& \xi^y_{i,t},\,\xi^u_{i,t},\,\xi^v_{i,t} \ge 0, \\
& z^y_{i,t},\,z^u_{i,t},\,z^v_{i,t} \in \{0,1\}.
\end{align}
\end{subequations}
Thus, (\ref{eq:sto_constraints}n) is enforced indirectly via binary proxies and vanishing slacks at convergence. $Z$ and $\Xi$ are nonanticipative, while $(p_{i,t,s},r^{\uparrow}_{i,t,s},r^{\downarrow}_{i,t,s})$ remain scenario dependent.

\subsubsection*{Augmented Lagrangian}

Let $\Lambda=\{\lambda^y,\lambda^u,\lambda^v\}$ be the multipliers, $\rho_y,\rho_u,\rho_v>0$ penalty parameters, and $\beta_y,\beta_u,\beta_v\ge 0$ proximal weights. Define the aggregate of all primal variables
\[
\Delta := \{\,y,u,v,p,r^{\uparrow},r^{\downarrow}\,\},
\]
with $(y,u,v)$ relaxed to $[0,1]$ in Block~1. The augmented Lagrangian is
\begin{align}\label{auglag}
\mathcal{L}(&\Delta,Z,\Xi,\Lambda) 
= \sum_{t,i} \big(A_i y_{i,t}+ S_i u_{i,t}+ H_i v_{i,t}\big) \nonumber\\
&+ \sum_{s} \pi_s \sum_{t,i}\big(B_i p_{i,t,s}+C_i p_{i,t,s}^2\big) \nonumber\\
&+ \sum_{i,t}\!\Big[\lambda^y_{i,t}(y_{i,t}-z^y_{i,t}+\xi^y_{i,t})
+\tfrac{\rho_y}{2}(y_{i,t}-z^y_{i,t}+\xi^y_{i,t})^2\Big] \nonumber\\
&+ \sum_{i,t}\!\Big[\lambda^u_{i,t}(u_{i,t}-z^u_{i,t}+\xi^u_{i,t})
+\tfrac{\rho_u}{2}(u_{i,t}-z^u_{i,t}+\xi^u_{i,t})^2\Big] \nonumber\\
&+ \sum_{i,t}\!\Big[\lambda^v_{i,t}(v_{i,t}-z^v_{i,t}+\xi^v_{i,t})
+\tfrac{\rho_v}{2}(v_{i,t}-z^v_{i,t}+\xi^v_{i,t})^2\Big] \nonumber\\
&+ \sum_{i,t}\!\Big[\tfrac{\beta_y}{2}(\xi^y_{i,t})^2
+ \tfrac{\beta_u}{2}(\xi^u_{i,t})^2
+ \tfrac{\beta_v}{2}(\xi^v_{i,t})^2\Big],
\end{align}
subject to (\ref{eq:sto_constraints}b)–(\ref{eq:sto_constraints}m).

\subsubsection*{ADMM Updates}

At iteration $k$, three primal blocks and a dual update are performed:

\paragraph{Block~1: Continuous update.}
\begin{align}
\Delta^{(k)} = \arg\min_{\Delta} \;
\mathcal{L}\big(\Delta, Z^{(k-1)}, \Xi^{(k-1)}, \Lambda^{(k-1)}\big),
\end{align}
subject to (\ref{eq:sto_constraints}b)–(\ref{eq:sto_constraints}m) with the relaxation $(y,u,v)\in[0,1]$. 
This is a convex QP in the relaxed $(y,u,v)$ and the scenario–dependent $(p,r^{\uparrow},r^{\downarrow})$.

\paragraph{Block~2: Binary update (QUBO).}
\begin{align}\label{zqubo}
Z^{(k)} = \arg\min_{Z\in\{0,1\}^{3NT}}
\mathcal{L}\big(\Delta^{(k)},Z,\Xi^{(k-1)},\Lambda^{(k-1)}\big).
\end{align}
This decouples into independent 1-bit QUBOs:
\begin{align}
q^{(y)}_{i,t} &= -\lambda^{y,(k-1)}_{i,t} - \rho_y\!\big(y^{(k)}_{i,t}+\xi^{y,(k-1)}_{i,t}\big) + \tfrac{\rho_y}{2},\label{eq:update_y} \\
q^{(u)}_{i,t} &= -\lambda^{u,(k-1)}_{i,t} - \rho_u\!\big(u^{(k)}_{i,t}+\xi^{u,(k-1)}_{i,t}\big) + \tfrac{\rho_u}{2},\label{eq:update_u} \\
q^{(v)}_{i,t} &= -\lambda^{v,(k-1)}_{i,t} - \rho_v\!\big(v^{(k)}_{i,t}+\xi^{v,(k-1)}_{i,t}\big) + \tfrac{\rho_v}{2},\label{eq:update_v}
\end{align}
with $z^\bullet_{i,t}\in \arg\min_{z\in\{0,1\}} q^{(\bullet)}_{i,t}\,z$. These 1-bit baselines are later extended to multi-bit micro-/batched QUBOs (Section~\ref{sec:qubo_dvqe}).

\paragraph{Block~3: Slack update.}
\begin{align}
\xi^{y,(k)}_{i,t} &= -\frac{\lambda^{y,(k-1)}_{i,t} + \rho_y\!\big(y^{(k)}_{i,t}-z^{y,(k)}_{i,t}\big)}{\beta_y+\rho_y}, 
\label{eq:slack_y} \\[4pt]
\xi^{u,(k)}_{i,t} &= -\frac{\lambda^{u,(k-1)}_{i,t} + \rho_u\!\big(u^{(k)}_{i,t}-z^{u,(k)}_{i,t}\big)}{\beta_u+\rho_u}, 
\label{eq:slack_u} \\[4pt]
\xi^{v,(k)}_{i,t} &= -\frac{\lambda^{v,(k-1)}_{i,t} + \rho_v\!\big(v^{(k)}_{i,t}-z^{v,(k)}_{i,t}\big)}{\beta_v+\rho_v},
\label{eq:slack_v}
\end{align}
followed by projection $\xi^\bullet_{i,t}\leftarrow \max\{\xi^\bullet_{i,t},0\}$.

\paragraph{Dual update.}
\begin{align}
\lambda^{y,(k)}_{i,t} &= \lambda^{y,(k-1)}_{i,t} + \rho_y\!\big(y^{(k)}_{i,t}-z^{y,(k)}_{i,t}+\xi^{y,(k)}_{i,t}\big),
\label{eq:dual_y} \\[4pt]
\lambda^{u,(k)}_{i,t} &= \lambda^{u,(k-1)}_{i,t} + \rho_u\!\big(u^{(k)}_{i,t}-z^{u,(k)}_{i,t}+\xi^{u,(k)}_{i,t}\big),
\label{eq:dual_u} \\[4pt]
\lambda^{v,(k)}_{i,t} &= \lambda^{v,(k-1)}_{i,t} + \rho_v\!\big(v^{(k)}_{i,t}-z^{v,(k)}_{i,t}+\xi^{v,(k)}_{i,t}\big).
\label{eq:dual_v}
\end{align}

\begin{algorithm}[t!]
  \caption{Three-Block ADMM for Stochastic UC}\label{alg:3b_admm}
  \begin{algorithmic}[1]
    \STATE \textbf{Initialize:} $(Z^{(0)}, \Xi^{(0)}, \Lambda^{(0)})$, penalty parameters, proximal weights, and tolerances. Set $k \gets 1$.
    \WHILE{stopping criterion not met}
      \STATE \textbf{Block 1:} 
      Solve $\Delta^{(k)}$ by minimizing \eqref{auglag} subject to (\ref{eq:sto_constraints}b)–(\ref{eq:sto_constraints}m) with $(y,u,v)\in[0,1]$.
      \STATE \textbf{Block 2:} 
      Update $Z^{(k)}$ by solving the QUBO subproblems defined in \eqref{zqubo}.
      \STATE \textbf{Block 3:} 
      Update $\Xi^{(k)}$ using \eqref{eq:slack_y}–\eqref{eq:slack_v} and project $\Xi^{(k)}\leftarrow\max\{\Xi^{(k)},0\}$.
      \STATE \textbf{Dual:} 
      Update $\Lambda^{(k)}$ via \eqref{eq:dual_y}–\eqref{eq:dual_v}.
      \STATE $k \gets k+1$.
    \ENDWHILE
    \STATE \textbf{Return:} $(\Delta^{(k)}, Z^{(k)}, \Xi^{(k)}, \Lambda^{(k)})$.
  \end{algorithmic}
\end{algorithm}

\paragraph*{Stopping rule.}
We terminate when the primal residuals $\|y-z^y+\xi^y\|_2$, $\|u-z^u+\xi^u\|_2$, $\|v-z^v+\xi^v\|_2$ and the dual residual $\rho\|(Z^{(k)}\!-\!Z^{(k-1)})-(\Xi^{(k)}\!-\!\Xi^{(k-1)})\|_2$ fall below thresholds $\varepsilon_{\text{pri}},\varepsilon_{\text{dual}}$, or when a maximum iteration cap is reached.

\section{Three-Block ADMM Convergence Analysis}\label{sec:Three_block_ADMM_Con}

We analyze the convergence of the three-block ADMM. We first state the conditions under which the method converges to a stationary solution, and then study stability via a Lyapunov-based argument.

\subsection{Convergence Analysis}

The three-block ADMM reformulation introduces consensus constraints that link the relaxed first-stage variables \((y,u,v)\in[0,1]^{N\times T}\) to binary proxies \((z^y,z^u,z^v)\in\{0,1\}^{N\times T}\) through nonnegative proximal slacks \((\xi^y,\xi^u,\xi^v)\ge 0\). At convergence, feasibility is recovered if the primal residuals
\begin{align}
&\|y - z^y + \xi^y\|_2 \to 0,\qquad \|u - z^u + \xi^u\|_2 \to 0, \nonumber\\
&\|v - z^v + \xi^v\|_2 \to 0,
\end{align}
and the slacks diminish elementwise, implying \(y=z^y\), \(u=z^u\), \(v=z^v\) with binary values.
Since the continuous variables \((p,r^{\uparrow},r^{\downarrow})\) remain within the feasible region
defined by (\ref{eq:sto_constraints}b)--(\ref{eq:sto_constraints}m), the limit points correspond to valid UC schedules with no relaxation gap.

\paragraph*{Sufficient conditions.}
The three-block ADMM iterations converge to a stationary solution of the UC problem under the following conditions \cite{gambella2020multiblock}:
\begin{enumerate}
\item \textbf{Coercivity.}  
The stochastic cost \eqref{eq:sto_obj} is coercive, since it contains convex quadratic terms \((C_i p_{i,t,s}^2)\) and all decision variables are bounded by capacity, ramping, and reserve constraints.  
Thus, the augmented Lagrangian \(\mathcal{L}\) is bounded.

\item \textbf{Feasibility (range condition).}  
The reformulated UC satisfies
\[
\operatorname{Im}(y,u,v) \subseteq \operatorname{Im}(z^y,z^u,z^v),
\]
because every relaxed variable has a corresponding proxy and slack.  
Therefore, primal feasibility is attainable.

\item \textbf{Lipschitz continuity.}  
The updates in Block~1 and Block~3 are Lipschitz continuous with a constant \(M=1\). Block~2 is a finite QUBO minimization and its value mapping over \(\{0,1\}\) is trivially Lipschitz.

\item \textbf{Prox-regularity and lower semicontinuity.}  
The UC cost is a sum of convex terms 
\((A_i y_{i,t},\,B_i p_{i,t,s},\,C_i p_{i,t,s}^2,\,S_i u_{i,t},\,H_i v_{i,t})\) 
and quadratic penalties in \(\mathcal{L}\), which are 
Lipschitz differentiable and prox-regular.  
Hence, \(\mathcal{L}\) is lower semicontinuous.

\item \textbf{Kurdyka--\L{}ojasiewicz (K\L{}) property.}  
Since the UC problem is defined by linear and quadratic relations with binary variables, it is semialgebraic. Therefore, \(\mathcal{L}\) satisfies the K\L{} inequality, ensuring the convergence of ADMM iterates to a stationary point.
\end{enumerate}

\medskip
Putting these together, if each block is solved and the penalty parameters \(\rho_y,\rho_u,\rho_v\) are chosen large, the three-block ADMM updates for stochastic UC converge to a stationary point of the original mixed-integer formulation.

\subsection{Stability via Lyapunov Analysis}

We assume throughout: 
(A1) the UC cost is coercive and all variables are bounded by (\ref{eq:sto_constraints}b)--(\ref{eq:sto_constraints}m); 
(A2) for fixed \((Z,S,\Lambda)\) the Block~1 subproblem admits a minimizer and is strongly convex in its local variables; 
(A3) Block~2 is solved.

\begin{proof}
Define the consensus residuals at iteration \(k\):
\begin{align}
&r^{y,k}:=y^{(k)}-z^{y,(k)}+\xi^{y,(k)}, 
r^{u,k}:=u^{(k)}-z^{u,(k)}+\xi^{u,(k)},\nonumber\\
&r^{v,k}:=v^{(k)}-z^{v,(k)}+\xi^{v,(k)}, 
\end{align}
and stack them as \(r^k:=(r^{y,k},r^{u,k},r^{v,k})\). Let \(\rho:=\mathrm{diag}(\rho_y I,\rho_u I,\rho_v I)\) and define the aggregated dual vector \(\lambda^k:=(\lambda^{y,(k)},\lambda^{u,(k)},\lambda^{v,(k)})\). The dual updates give
\begin{equation}\label{eq:dual_update_compact}
\lambda^{k+1}=\lambda^{k}+\rho\,r^{k+1}\quad\Longrightarrow\quad
r^{k+1}=\rho^{-1}\!\left(\lambda^{k+1}-\lambda^k\right).
\end{equation}
Consider the Lyapunov candidate
\begin{align}\label{eq:V_def}
V^k
:=\mathcal{L}\!\big(\Delta^{(k)},Z^{(k)},\Xi^{(k)},\Lambda^{(k)}\big)
+\frac{\kappa}{2}\,\|r^k\|_2^2,
\end{align}
where \(\kappa\ge 0\). We show \(V^{k+1}-V^k\le 0\) and deduce convergence.

\paragraph*{Step 1: Lyapunov difference expansion.}
\begin{align}\label{labela}
V^{k+1}-V^k
=&\underbrace{\Big(\mathcal{L}^{k+1}(\lambda^{k+1})-\mathcal{L}^{k}(\lambda^{k})\Big)}_{\mathcal{T}_\mathcal{L}}
\nonumber\\&+\frac{\kappa}{2}\big(\|r^{k+1}\|^2-\|r^{k}\|^2\big).
\end{align}

\paragraph*{Step 2: Augmented Lagrangian change.}
Split \(\mathcal{T}_{\mathcal{L}}\) into a dual-shift part (changing \(\Lambda\) with fixed
primals) and block updates (fixed \(\Lambda\)):
\begin{align}
\mathcal{T}_{\mathcal{L}}
=& \big[\mathcal{L}^{k+1}(\lambda^{k+1})-\mathcal{L}^{k+1}(\lambda^{k})\big]
 + \big[\mathcal{L}^{k+1}(\lambda^{k})-\mathcal{L}^{k}(\lambda^{k})\big] \nonumber\\
=& \langle \lambda^{k+1}\!-\!\lambda^{k},\, r^{k+1}\rangle
 + \Big(
 \mathcal{L}\!\big(\Delta^{(k+1)},Z^{(k+1)},\Xi^{(k+1)},\Lambda^{(k)}\big)
 \nonumber\\&-\mathcal{L}\!\big(\Delta^{(k)},Z^{(k)},\Xi^{(k)},\Lambda^{(k)}\big)
 \Big).\label{eq:T_L_split}
\end{align}
Using \eqref{eq:dual_update_compact}, the dual-shift term equals
\(\langle \rho r^{k+1},\,r^{k+1}\rangle=\|\rho^{1/2}r^{k+1}\|^2\).

\paragraph*{Step 3: Descent by blocks at fixed duals.}
By (A2) and optimality of Block~1,
\begin{align}
\mathcal{L}\big(\Delta^{(k+1)},Z^{(k)},\Xi^{(k)},\Lambda^{(k)}\big)
&\le
\mathcal{L}\big(\Delta^{(k)},Z^{(k)},\Xi^{(k)},\Lambda^{(k)}\big)
\nonumber\\&-\sigma_1\|\Delta^{(k+1)}-\Delta^{(k)}\|^2.  
\end{align}
By (A3) for Block~2,
\begin{align}
 \mathcal{L}\big(\Delta^{(k+1)},Z^{(k+1)},\Xi^{(k)},\Lambda^{(k)}\big)
\le
\mathcal{L}\big(\Delta^{(k+1)},Z^{(k)},\Xi^{(k)},\Lambda^{(k)}\big).   
\end{align}
For Block~3, the proximal quadratic update (with projection) yields
\begin{align}
\mathcal{L}\big(&\Delta^{(k+1)},Z^{(k+1)},\Xi^{(k+1)},\Lambda^{(k)}\big)
\le
\nonumber\\&\mathcal{L}\big(\Delta^{(k+1)},Z^{(k+1)},\Xi^{(k)},\Lambda^{(k)}\big)
-\sigma_2\|\Xi^{(k+1)}-\Xi^{(k)}\|^2.
\label{eq:fixed_dual_descent}
\end{align}

\paragraph*{Step 4: Collect terms.}
Substituting \eqref{eq:T_L_split} and \eqref{eq:fixed_dual_descent} into \eqref{labela}:
\begin{align}
V^{k+1}-V^k
&\le
\|\rho^{1/2}r^{k+1}\|^2
-\sigma_1\|\Delta^{(k+1)}-\Delta^{(k)}\|^2
\nonumber\\
&\quad-\sigma_2\|\Xi^{(k+1)}-\Xi^{(k)}\|^2
+\frac{\kappa}{2}\big(\|r^{k+1}\|^2-\|r^{k}\|^2\big)
\nonumber\\&\le
\Big(\lambda_{\max}(\rho)+\tfrac{\kappa}{2}\Big)\|r^{k+1}\|^2
-\tfrac{\kappa}{2}\|r^{k}\|^2
\nonumber\\&\quad
-\sigma_1\|\Delta^{(k+1)}-\Delta^{(k)}\|^2
-\sigma_2\|\Xi^{(k+1)}-\Xi^{(k)}\|^2.
\label{eq:Vdiff_final}
\end{align}

\paragraph*{Step 5: Choice of \(\kappa\) and descent.}
Choose \(\kappa\ge 2\,\lambda_{\max}(\rho)\) so that
\(\lambda_{\max}(\rho)+\frac{\kappa}{2}\le \kappa\).
Then \eqref{eq:Vdiff_final} becomes
\begin{align}\label{eq:lyapunov_final}
 V^{k+1}-V^k
&\le
\kappa\|r^{k+1}\|^2-\frac{\kappa}{2}\|r^{k}\|^2
-\sigma_1\|\Delta^{(k+1)}-\Delta^{(k)}\|^2
\nonumber\\&\quad-\sigma_2\|\Xi^{(k+1)}-\Xi^{(k)}\|^2.   
\end{align}
Summing over \(k\) and using that \(V^k\) is bounded below (coercivity, (A1)) yields
\[
\|\Delta^{(k+1)}-\Delta^{(k)}\|\to 0,\quad
\|\Xi^{(k+1)}-\Xi^{(k)}\|\to 0,\quad
\|r^{k}\|\to 0,
\]
by a standard telescoping argument. Hence \(V^k\) is non-increasing, and the iterates converge, with binary feasibility recovered via vanishing slacks.
\end{proof}

\section{QUBO Decompositions and DVQE Integration}\label{sec:qubo_dvqe}
We formulate the Block 2 binary update as a quadratic unconstrained binary optimization (QUBO) and develop three decomposition strategies—(i) type-specific (commitment/startup/shutdown), (ii) micro (per-unit, per-time), and (iii) batched block-diagonal—to create sub-QUBOs. We then introduce a distributed VQE solver (DVQE) with an accept-if-better safeguard to execute these sub-QUBOs efficiently across multiple quantum devices in a distributed quantum–classical workflow.

Within the three-block ADMM structure, the discrete subproblem (Block~2) isolates the auxiliary binary proxies $Z=\{z^y,z^u,z^v\}$, while all continuous and relaxed variables are handled in Block~1 and consensus is enforced in Block~3. At iteration $k$, the coefficients of $Z$ are as \eqref{eq:update_y}--\eqref{eq:update_v} with $z^\bullet_{i,t} \in \arg\min_{z\in\{0,1\}} q^{(\bullet)}_{i,t}z$.  

Although the QUBO formulation is straightforward, it presents two main challenges. The first is that the binary variables of each unit are not independent. Startup, shutdown, and commitment decisions are logically coupled through minimum up/down time conditions and other temporal constraints. These couplings are not directly visible in the linear objective above because they have been shifted into Block~1 and the consensus conditions. To ensure consistency of UC schedules, Block~2 must eventually include additional quadratic penalties that reintroduce these local dependencies. The second challenge is scalability. For large-scale UC problems, the monolithic QUBO corresponding to Block~2 can contain thousands of binary variables.

A further consideration is the target hardware environment. The purpose of isolating Block~2 is to cast UC in binary form and make it compatible with hybrid quantum–classical algorithms. To be effective, the QUBO must be large enough to demonstrate a potential advantage of quantum solvers over purely classical methods, but not so large that it overwhelms the limited qubits and circuit depth of current processors. Balancing these requirements motivates the decomposition strategies developed in the following subsections.

\subsection{QUBO Formulation}\label{sec:monolithic_qubo}

\subsubsection{Three QUBOs}\label{sec:three_qubos}

The initial form of Block~2 shows that all binary proxies appear independently in the augmented Lagrangian. This independence motivates a first decomposition strategy in which QUBO is partitioned into three smaller subproblems, each corresponding to a distinct type of binary variable. Specifically, all auxiliary commitment binaries $\{z^y_{i,t}\}$ are grouped into one QUBO, all startup binaries $\{z^u_{i,t}\}$ form a second QUBO, and all shutdown binaries $\{z^v_{i,t}\}$ form a third QUBO. In this way, the burden of solving a very large QUBO in one attempt is avoided, while no physical or mathematical structure of the UC model is violated. The three sub-QUBOs remain fully consistent with the consensus framework, and together, they constitute the complete Block~2 update.

Formally, the decomposition yields
\begin{align}
Z^{(k)} 
&= \arg\min_{Z\in\{0,1\}^{3NT}}
\mathcal{L}\big(\Delta^{(k)},Z,\Xi^{(k-1)},\Lambda^{(k-1)}\big) \\
&= 
\arg\min_{\substack{z^y\in\{0,1\}^{NT}\\ z^u\in\{0,1\}^{NT}\\ z^v\in\{0,1\}^{NT}}}
\Big(\mathcal{L}_y(z^y) + \mathcal{L}_u(z^u) + \mathcal{L}_v(z^v)\Big),
\end{align}
where $\mathcal{L}_y$, $\mathcal{L}_u$, and $\mathcal{L}_v$ denote the components of the augmented Lagrangian 
associated with the three classes of binary variables. 
Each subproblem is thus a self-contained QUBO with a linear objective over its respective binary set. 

This decomposition provides two main benefits. First, the size of each QUBO is reduced by roughly a factor of three, which lowers the computational complexity of both classical and hybrid solvers. Second, the three QUBOs can be solved in parallel, thereby preserving the interpretation of Block~2 as a single update while reducing wall-clock time per ADMM iteration. 

\subsubsection{Micro QUBOs}\label{sec:micro_qubos}

The independent decomposition into three type-specific QUBOs reduces problem size but does not restore the logical dependencies that exist among commitment, startup, and shutdown variables of the same unit. To better capture these local couplings while maintaining the tractability of Block~2, we introduce a micro-level decomposition. The guiding idea is to reconsider, at the level of auxiliary proxies, the same relations that the relaxed variables $(y,u,v)$ must satisfy in Block~1. Since these dependencies occur only within each unit individually, a natural approach is to construct one micro-QUBO per unit–time pair.

At each $(i,t)$ we bundle the three proxies $z^y_{i,t},z^u_{i,t},z^v_{i,t}\in\{0,1\}$ into a three-dimensional binary vector.
\[
z_{i,t} := \begin{bmatrix} z^y_{i,t} & z^u_{i,t} & z^v_{i,t} \end{bmatrix} \in \{0,1\}^3,
c_{i,t} := \begin{bmatrix} q^{(y)}_{i,t} & q^{(u)}_{i,t} & q^{(v)}_{i,t} \end{bmatrix},
\]
where the entries of $c_{i,t}$ are the linear coefficients inherited from the augmented Lagrangian. We then define a local quadratic energy
\begin{equation}
E_{i,t}(z_{i,t}) \;=\; z_{i,t}^\top Q_{i,t}\, z_{i,t} \;+\; c_{i,t}^\top z_{i,t} \;+\; \text{const},
\label{eq:micro_qubo_energy}
\end{equation}
with $Q_{i,t}\in\mathbb{R}^{3\times 3}$ symmetric. In contrast to the type-specific decomposition, here we deliberately augment the base linear terms with small quadratic penalties that reintroduce period-$t$ logic. These penalties are purely local and maintain independence across different units, allowing all microproblems to be solved in parallel.

To construct $Q_{i,t}$, we let the relaxed Block~1 outputs at $(i,t)$ be $\hat y:=y^{(k)}_{i,t}$, $\hat u:=u^{(k)}_{i,t}$, and $\hat v:=v^{(k)}_{i,t}$, with a reference $\eta:=y_{t-1}^{\mathrm{ref}}$ from the previous period. With nonnegative penalty weights $\gamma_c,\gamma_{ss},\gamma_{u\to y},\gamma_{v\to\bar y},\gamma_y,\gamma_u,\gamma_v$, we define the penalty
\begin{align}
\Pi_{i,t}(z_{i,t}) :=&
\gamma_c\big(z^y_{i,t} - z^u_{i,t} + z^v_{i,t} - \eta\big)^2 + \gamma_{ss} z^u_{i,t} z^v_{i,t}
\nonumber\\
&+ \gamma_{u\to y} z^u_{i,t}(1-z^y_{i,t})
+ \gamma_{v\to\bar y} z^v_{i,t} z^y_{i,t} \nonumber\\
&+ \gamma_y (z^y_{i,t}-\hat y)^2
+ \gamma_u (z^u_{i,t}-\hat u)^2
+ \gamma_v (z^v_{i,t}-\hat v)^2.
\label{eq:micro_penalties}
\end{align}
The first term softly enforces the logical relation (\ref{eq:sto_constraints}d). The second discourages simultaneous startup and shutdown. The third and fourth encode start implies on and shutdown implies off. The last three anchor the proxies to the relaxed outputs from Block~1. These additions are incorporated by setting $E_{i,t}\gets E_{i,t}+\Pi_{i,t}$.

Because binary variables satisfy $z^2=z$, the penalty \eqref{eq:micro_penalties} contributes explicit updates to $(Q_{i,t},c_{i,t})$. The linear terms are updated as
\begin{align}
c^{(y)}_{i,t} &\leftarrow c^{(y)}_{i,t}+\gamma_c(1-2\eta)+\gamma_y(1-2\hat y), \\
c^{(u)}_{i,t} &\leftarrow c^{(u)}_{i,t}+\gamma_c(1+2\eta)+\gamma_u(1-2\hat u)+\gamma_{u\to y}, \\
c^{(v)}_{i,t} &\leftarrow c^{(v)}_{i,t}+\gamma_c(1-2\eta)+\gamma_v(1-2\hat v),
\end{align}
while the pairwise couplings are
\begin{align}
(Q_{i,t})_{y,u} &\leftarrow (Q_{i,t})_{y,u}-2\gamma_c-\gamma_{u\to y}, \\
(Q_{i,t})_{y,v} &\leftarrow (Q_{i,t})_{y,v}+2\gamma_c+\gamma_{v\to\bar y}, \\
(Q_{i,t})_{u,v} &\leftarrow (Q_{i,t})_{u,v}-2\gamma_c+\gamma_{ss}.
\end{align}
We initialize $Q_{i,t}=0_{3\times 3}$ and $c_{i,t}=[q^{(y)}_{i,t},q^{(u)}_{i,t},q^{(v)}_{i,t}]^\top$, then apply these updates. The Block~2 update thus decomposes into independent micro problems
\begin{equation}
z^{\star}_{i,t}=\arg\min_{z_{i,t}\in\{0,1\}^3}\; z_{i,t}^\top Q_{i,t} z_{i,t}+c_{i,t}^\top z_{i,t}.
\label{eq:micro_qubo}
\end{equation}
This approach restores local binary couplings and accelerates ADMM convergence.

\subsubsection{Batched QUBOs}\label{sec:grouped_qubos}

The micro-QUBO decomposition accelerates ADMM by introducing local couplings and reducing the size of the Block~2 subproblems compared to the three independent QUBOs. However, this decomposition alone does not exploit quantum efficiency. Each micro-QUBO involves only three binary variables, so even classical brute force can enumerate its eight possible assignments in negligible time. Moreover, in large-scale UC instances with $N$ units, $T$ periods, and multiple ADMM iterations, the total number of micro problems becomes $3NT\times(\text{iterations})$, which may reach millions. Solving such a volume of micro problems in parallel quickly exceeds computational resources, while solving them sequentially becomes prohibitively time-consuming. 

The remedy lies in re-aggregating the micro problems into carefully chosen batches. Instead of solving all $NT$ micro-QUBOs independently, we group them into $k$ disjoint batches, with $3<k<N$. Each batch is large enough to justify the use of quantum solvers but small enough to remain within the current qubit limitations of near-term quantum hardware. In addition, the grouping allows us to balance solver effort by assigning harder micro problems to different batches. A greedy distribution algorithm ensures that batches have roughly equal difficulty, allowing all workers to complete in comparable times. This approach thus interpolates between the two earlier decompositions: it yields more subproblems than the three-QUBO strategy, while preserving the local couplings absent in the fully independent decomposition, and it reduces the computational overhead of the micro-QUBO strategy by replacing millions of subproblems.

Although each micro-QUBO has only three variables, its solver difficulty depends on the local energy landscape of the problem. To balance batches, we compute a simple hardness score $\mathrm{Hard}_{i,t}$ from $(Q_{i,t},c_{i,t})$. Several complementary indicators are combined:
\begin{align}
 \mathrm{Hard}_{i,t}
= &\; w_1\,\frac{1}{\Delta_{i,t}+\eta}
+ w_2\,\frac{g^{(\varepsilon)}_{i,t}-1}{7}
+ w_3\,\mathbb{I}_{\mathrm{frust},i,t} \nonumber\\
&+ w_4\,\frac{r_{i,t}}{1+r_{i,t}}
+ w_5\,\frac{\log_{10}(M_{i,t}/m_{i,t})}{4},
\end{align}
where the terms are:
\begin{itemize}
\item $\Delta_{i,t}$: the energy gap between the best and second-best assignment (larger gap = easier);
\item $g^{(\varepsilon)}_{i,t}$: number of assignments within $\varepsilon$ of the optimum (larger = harder);
\item $\mathbb{I}_{\mathrm{frust},i,t}$: frustration indicator for the 3-variable coupling triangle;
\item $r_{i,t}$: ratio of total coupling strength to total linear field strength;
\item $\log_{10}(M_{i,t}/m_{i,t})$: dynamic range of coefficients, with $M_{i,t}$ and $m_{i,t}$ the maximum and minimum nonzero absolute entries of $Q_{i,t}$ and $c_{i,t}$.
\end{itemize}
The weights $w_1,\dots,w_5$ are user-tunable (initially equal), and $\eta>0$ avoids division by zero. This score is inexpensive to compute and serves as a proxy for solver effort.

Given scores $\mathrm{Hard}_{i,t}$, we partition the set $\{(i,t)\}$ into $k$ disjoint batches $B_1,\dots,B_k$ with similar total hardness. A simple greedy bin-packing procedure suffices. Firstly, sort all pairs $(i,t)$ by decreasing $\mathrm{Hard}_{i,t}$ and secondly, assign each next element to the batch whose current hardness sum is smallest.

This balances the workload across batches. As a practical refinement, one may enforce that all time periods of the same unit $(i,1),\dots,(i,T)$ are assigned to the same batch, preserving temporal coherence without overlap. Since batches are non-overlapping and each micro-QUBO couples only its own triplet $(z^y_{i,t},z^u_{i,t},z^v_{i,t})$, batching preserves the independence of subproblems. Thus, ADMM convergence is unaffected.

\subsection{DVQE Solver}\label{sec:dvqe}

As discussed earlier, the size of monolithic or even moderately decomposed QUBOs can exceed the qubit capacities of current NISQ devices. While UC admits natural decompositions across units and time, many other QUBO formulations are not inherently separable, which further exacerbates scalability challenges. As problem sizes grow, relying on a single QPU becomes infeasible. A promising direction to overcome this barrier is distributed quantum computing, where multiple smaller QPUs collaborate to solve a larger optimization task. The DVQE \cite{hasanzadeh2025distributed} is one such hybrid quantum–classical framework. DVQE extends the standard VQE algorithm by partitioning a QUBO across multiple quantum devices. Each QPU evaluates a subset of parameterized quantum circuits, while a classical coordination layer aggregates partial results, updates the global variational parameters, and drives the optimization forward. The TeleGate protocol enables entangling operations between qubits on different devices, creating the illusion of a single larger QPU.  

When working with DVQE, the user only provides the quadratic and linear coefficients of the QUBO. Internally, these coefficients are mapped into an Ising Hamiltonian written in terms of Pauli-$Z$ operators as
\begin{equation}
    H = \sum_{i=1}^{n} h_i Z_i + \sum_{i<j} J_{ij} Z_i Z_j ,
    \label{eq:ising}
\end{equation}
where $Z_i$ is the Pauli-$Z$ operator acting on qubit $i$, $h_i$ is the linear field, and $J_{ij}$ encodes pairwise couplings. The ground state of this Hamiltonian corresponds to the optimal QUBO solution. This automatic transformation allows DVQE to remain solver-agnostic with respect to the problem input.  

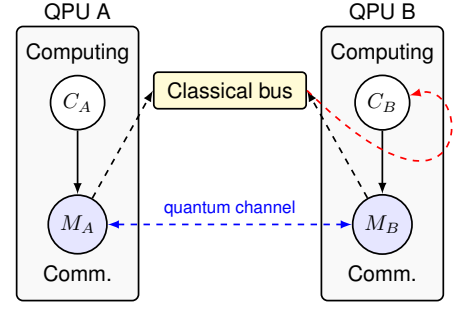
\begin{figure}[!t]
\centering
\resizebox{0.7\columnwidth}{!}{%
\begin{tikzpicture}[font=\sffamily, thick, >=latex]

\tikzset{
  qpu/.style = {rectangle, rounded corners=3pt, draw=black, thick, minimum width=3.0cm, minimum height=4.5cm, fill=gray!5},
  qubit/.style = {circle, draw=black, thick, minimum size=8mm, fill=white},
  comm/.style  = {circle, draw=black, thick, minimum size=8mm, fill=blue!10},
  arrow/.style = {->, thick}
}

\node[qpu, minimum width=2.0cm] (qpuA) at (-2.5,0) {};
\node at (-2.5,2.5) {QPU A};  

\node[qpu, minimum width=2.0cm] (qpuB) at (2.5,0) {};
\node at (2.5,2.5) {QPU B};   

\node[qubit] (compA) at (-2.5,1.0) {$C_A$};
\node[comm]  (commA) at (-2.5,-1.0) {$M_A$};
\node at (-2.5,1.8) {Computing};
\node at (-2.5,-1.8) {Comm.};

\node[qubit] (compB) at (2.5,1.0) {$C_B$};
\node[comm]  (commB) at (2.5,-1.0) {$M_B$};
\node at (2.5,1.8) {Computing};
\node at (2.5,-1.8) {Comm.};

\draw[arrow] (compA) -- (commA);
\draw[arrow] (compB) -- (commB);

\draw[<->, thick, blue, dashed] (commA) -- node[above,sloped] {\footnotesize quantum channel} (commB);

\node[draw=black, fill=yellow!20, rounded corners=2pt, minimum width=2.5cm, minimum height=0.6cm] (bus) at (0,1.2) {Classical bus};

\draw[arrow, dashed] (commA) -- (bus.west);
\draw[arrow, dashed] (commB) -- (bus.east);

\draw[arrow, dashed, red] (bus.east) .. controls (4.0,-1.5) and (4.0,1.4) .. (compB);

\end{tikzpicture}
}
\caption{\footnotesize Illustration of the TeleGate protocol.}
\label{fig:telegate}
\end{figure}

After the Hamiltonian is formed, qubits are distributed across QPUs using a greedy allocation algorithm that balances computational load while respecting the coupling structure. Each QPU hosts computing qubits for local gates and reserves at least one communication qubit that is used to entangle with other devices. Local operations such as single-qubit rotations are executed directly within a QPU, while inter-QPU entanglement is enabled by the TeleGate protocol. TeleGate combines local entanglement, quantum communication, and classical coordination. A computing qubit is first entangled with its local communication qubit. Communication qubits of two different QPUs are then entangled through a shared channel. Finally, a classical message transmits the outcome, allowing the target QPU to apply the correct conditional operation. Through this process, several small QPUs and the classical bus collectively behave as if they were one larger processor, making it possible to run distributed ansatz circuits that preserve coherence across devices as shown in Fig. \ref{fig:telegate}. 

At the heart of VQE and DVQE lies the construction of an ansatz, a parameterized quantum circuit designed to approximate the ground state of the Hamiltonian. The ansatz prepares a quantum state $|\psi(\theta)\rangle$ using layers of single-qubit rotations, typically $R_y$ and $R_z$, interleaved with CNOT gates that introduce entanglement. By stacking multiple layers, the circuit gains expressive power to capture increasingly complex correlations. Once the state is prepared, repeated measurements provide an estimate of the Hamiltonian expectation value, which serves as the cost function to be minimized.  

The optimization loop is closed by a classical optimizer, which updates the variational parameters based on measurement outcomes. DVQE employs ADAM \cite{kingma2014adam}, which maintains exponentially weighted averages of gradients and squared gradients and uses bias correction to stabilize the updates. Compared with gradient-free optimizers, ADAM achieves faster convergence and greater robustness under noisy measurements, which is particularly valuable in distributed settings. The prepare–measure–update cycle continues until convergence or until a maximum number of iterations is reached. After convergence, the optimized ansatz is executed to produce a histogram of candidate bitstrings, from which the best-performing solution is selected as the binary vector.  

In practice, DVQE has been implemented in the \texttt{raiselab} package, which supports distributed circuit execution and incorporates advanced initialization strategies such as black hole, gray wolf, and artificial bee colony metaheuristics to accelerate convergence. Fig.~\ref{fig:vqe_flowchart} illustrates the general workflow of the DVQE algorithm, and Fig.~\ref{fig:dvqe_workflow} summarizes the DVQE process, including Hamiltonian generation, qubit distribution, initialization, training, and solution extraction.  

In principle, DVQE can solve the full Block~2 QUBO in one step; however, the implementation in \cite{hasanzadeh2025distributed} currently relies on a Qiskit simulator on classical CPUs, which limits scalability. On real distributed hardware, DVQE would naturally scale to larger UC instances. We validate DVQE on small cases and recommend its use on decomposed QUBOs for present-day feasibility. While micro-QUBOs are too small for quantum gain, DVQE is well-suited for batched QUBOs, where efficiency and scalability are both achieved.
\begin{figure}[!t]
    \centering
    \includegraphics[width=0.85\linewidth]{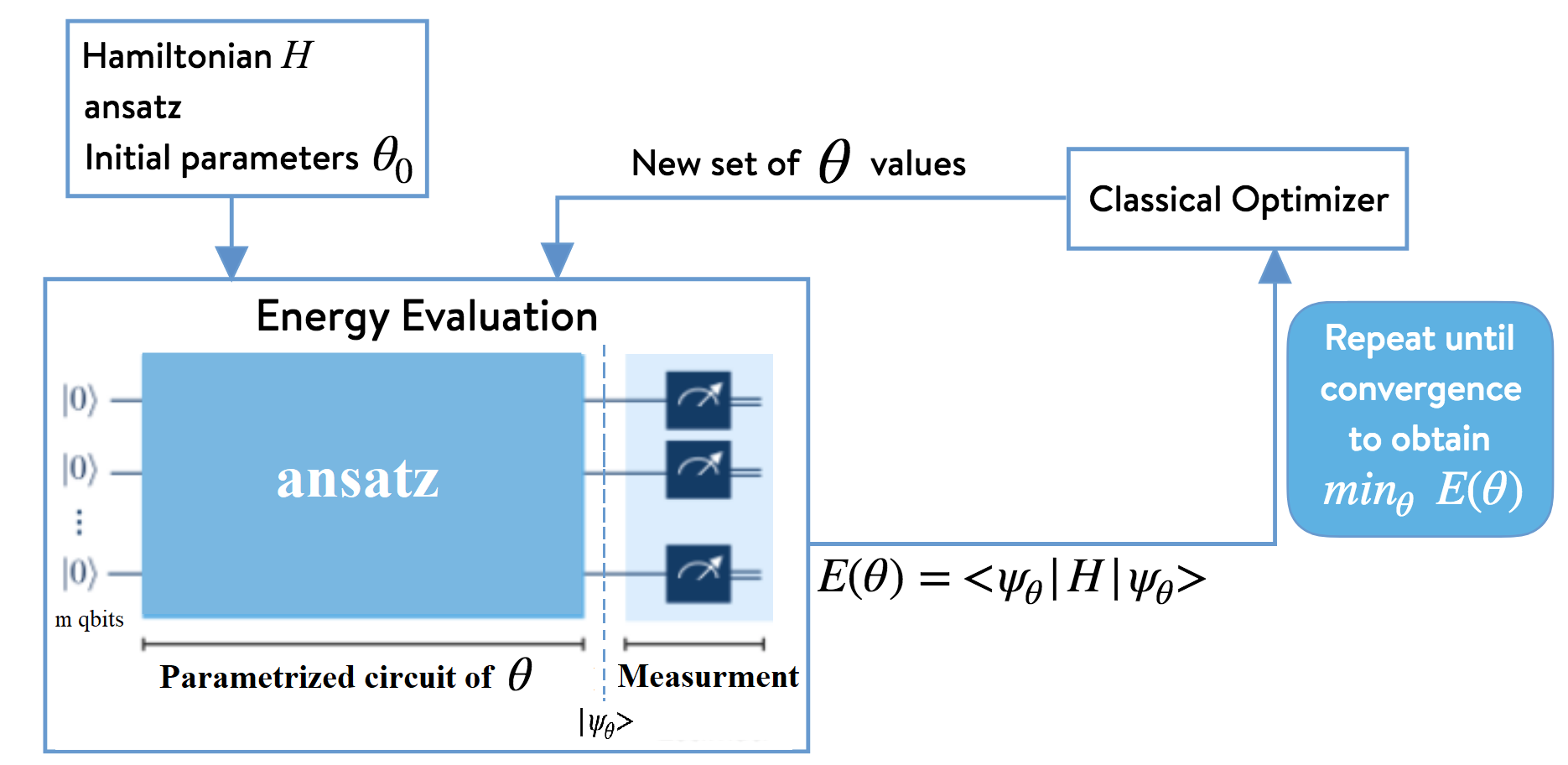}
    \caption{\footnotesize Overview of the VQE workflow for solving the Hamiltonian.}
    \label{fig:vqe_flowchart}
\end{figure}  
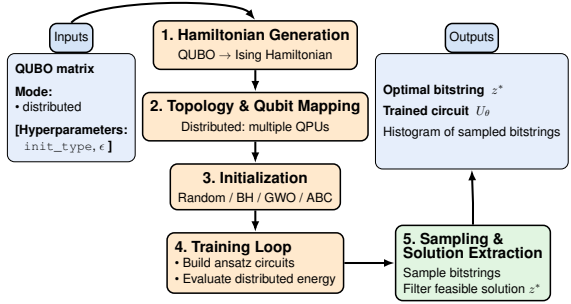
\begin{figure}[!t]
\centering
\resizebox{0.85\columnwidth}{!}{%
\begin{tikzpicture}[font=\sffamily]

\definecolor{headblue}{RGB}{213,228,247}
\definecolor{boxblue}{RGB}{230,238,251}
\definecolor{amberfill}{RGB}{255,233,205}
\definecolor{mintfill}{RGB}{225,245,225}

\tikzset{
  flowstep/.style   = {rounded corners=2mm, draw=black, very thick, fill=amberfill,
                       inner sep=6pt, minimum width=7.6cm, align=center},
  ioHeader/.style   = {rounded corners=2mm, draw=black, very thick, fill=headblue,
                       minimum width=5.4cm, minimum height=8mm, align=center},
  ioBody/.style     = {rounded corners=2mm, draw=black, line width=0.8pt, fill=boxblue,
                       inner sep=6pt, minimum width=5.4cm, align=left},
  samplebox/.style  = {rounded corners=2mm, draw=black, very thick, fill=mintfill,
                       inner sep=6pt, minimum width=6.0cm, align=left}
}

\node (inHeader) [ioHeader, minimum width=1.0cm] at (-5.1,3.7) {Inputs};
\node (inBody)   [ioBody, below=0pt of inHeader.south, anchor=north,
                  minimum height=3.0cm, minimum width=3.2cm] {%
\textbf{QUBO matrix}\\[2mm]
\textbf{Mode:}\\
\textbullet\  distributed\\[2mm]
\textbf{[Hyperparameters:}\\
\hspace*{0.7em}\texttt{init\_type}, $\epsilon$ \textbf{]}
};

\node (outHeader) [ioHeader, minimum width=1.0cm] at (6.0,3.7) {Outputs};
\node (outBody)   [ioBody, below=0pt of outHeader.south, anchor=north,
                   minimum height=3.2cm, minimum width=3.2cm] {%
\textbf{Optimal bitstring } $z^{*}$\\[1.5mm]
\textbf{Trained circuit } $U_{\theta}$\\[1.5mm]
Histogram of sampled bitstrings
};

\node (s1) [flowstep, minimum width=4.0cm] at (0,3.5) {%
{\large\bfseries 1.\ Hamiltonian Generation}\\[1mm]
QUBO $\to$ Ising Hamiltonian
};
\node (s2) [flowstep, below=6mm of s1, minimum width=4.0cm] {%
{\large\bfseries 2.\ Topology \& Qubit Mapping}\\[1mm]
Distributed: multiple QPUs
};
\node (s3) [flowstep, below=6mm of s2, minimum width=4.0cm] {%
{\large\bfseries 3.\ Initialization}\\[1mm]
Random / BH / GWO / ABC 
};
\node (s4) [flowstep, below=6mm of s3, align=left, minimum width=4.0cm] {%
{\large\bfseries 4.\ Training Loop}\\[-0.0mm]
\textbullet\ Build ansatz circuits\\
\textbullet\ Evaluate distributed energy
};

\node (s5) [samplebox, right=14mm of s4, minimum width=4.0cm] {%
{\large\bfseries 5.\ Sampling \&}\\[-0.5mm]
{\large\bfseries Solution Extraction}\\[1mm]
Sample bitstrings\\
Filter feasible solution $z^{*}$
};

\draw[-{Latex[length=3mm]}, line width=1.6pt]
  (inHeader.north) .. controls (-4.5,4.8) and (-0.5,4.8) .. (s1.north);

\draw[-{Latex[length=3mm]}, line width=1.6pt] (s1) -- (s2);
\draw[-{Latex[length=3mm]}, line width=1.6pt] (s2) -- (s3);
\draw[-{Latex[length=3mm]}, line width=1.6pt] (s3) -- (s4);

\draw[-{Latex[length=3mm]}, line width=1.6pt] (s4.east) -- (s5.west);

\draw[-{Latex[length=3mm]}, line width=1.6pt]
  (s5.north) -- ++(0,0.6) -- (outBody.south);

\end{tikzpicture}
}
\caption{\footnotesize Workflow of the DVQE algorithm.}
\label{fig:dvqe_workflow}
\end{figure}

\section{Case Studies and Results}\label{sec:results}

Across all D²-UC experiments, we fix DVQE hyperparameters for consistency: depth $d{=}2$, learning rate $\eta{=}0.1$, and at most 100 iterations. DVQE is run in distributed mode with disjoint 3-qubit registers (\texttt{[3,3,3,...,3]}). An accept-if-better safeguard ensures Block~2 updates monotonically reduce the local augmented Lagrangian, preventing oscillations. This configuration balances stability and runtime, and is used uniformly for the three-QUBO, micro-QUBO, and batched QUBO experiments. \ul{All detailed data, codes, and parameter values for all cases studied here with D²-UC framework are available at} \href{https://github.com/LSU-RAISE-LAB/3B-ADMM-UC-DVQE}{GitHub}.  

To ensure a fair comparison, we used identical ADMM settings and initialization in every case even for three-unit UC system: $\rho_y{=}\rho_u{=}\rho_v{=}9{\times}10^{5}$, $\beta_y{=}\beta_u{=}\beta_v{=}2{\times}10^{6}$, $\varepsilon{=}10^{-3}$, $\text{max\_iter}{=}4000$, micro-QUBO weights $\gamma_c{=}0.20\rho_y$, $\gamma_{ss}{=}0.10\rho_y$, $\gamma_{u\to y}{=}\gamma_{v\to\bar y}{=}0.05\rho_y$, and anchors $\gamma_y{=}\gamma_u{=}\gamma_v{=}0.10\rho_y$. Initialization follows: $p_0{=}\mathbf{20}$, $y_0{=}\mathbf{1}$, relaxed $(y,u,v)$ as $(0.5,0,0)$, $p$ tiled from $L/N$, zero slacks/duals, and zero reserves.

\subsection{Three-Unit UC}

We first validate the framework on a three-unit UC system with a planning horizon of $T{=}6$ hours. For this illustrative case, the full Block~2 QUBO is solved directly using DVQE without any decomposition, to demonstrate DVQE's ability to handle the binary optimization task as a whole. For benchmarking, the same problem is solved via exhaustive brute-force enumeration.  

\begin{figure}[!t]
    \captionsetup{font={footnotesize}}
    \centering
    \begin{subfigure}[b]{0.48\linewidth}
        \centering
        \includegraphics[width=\linewidth]{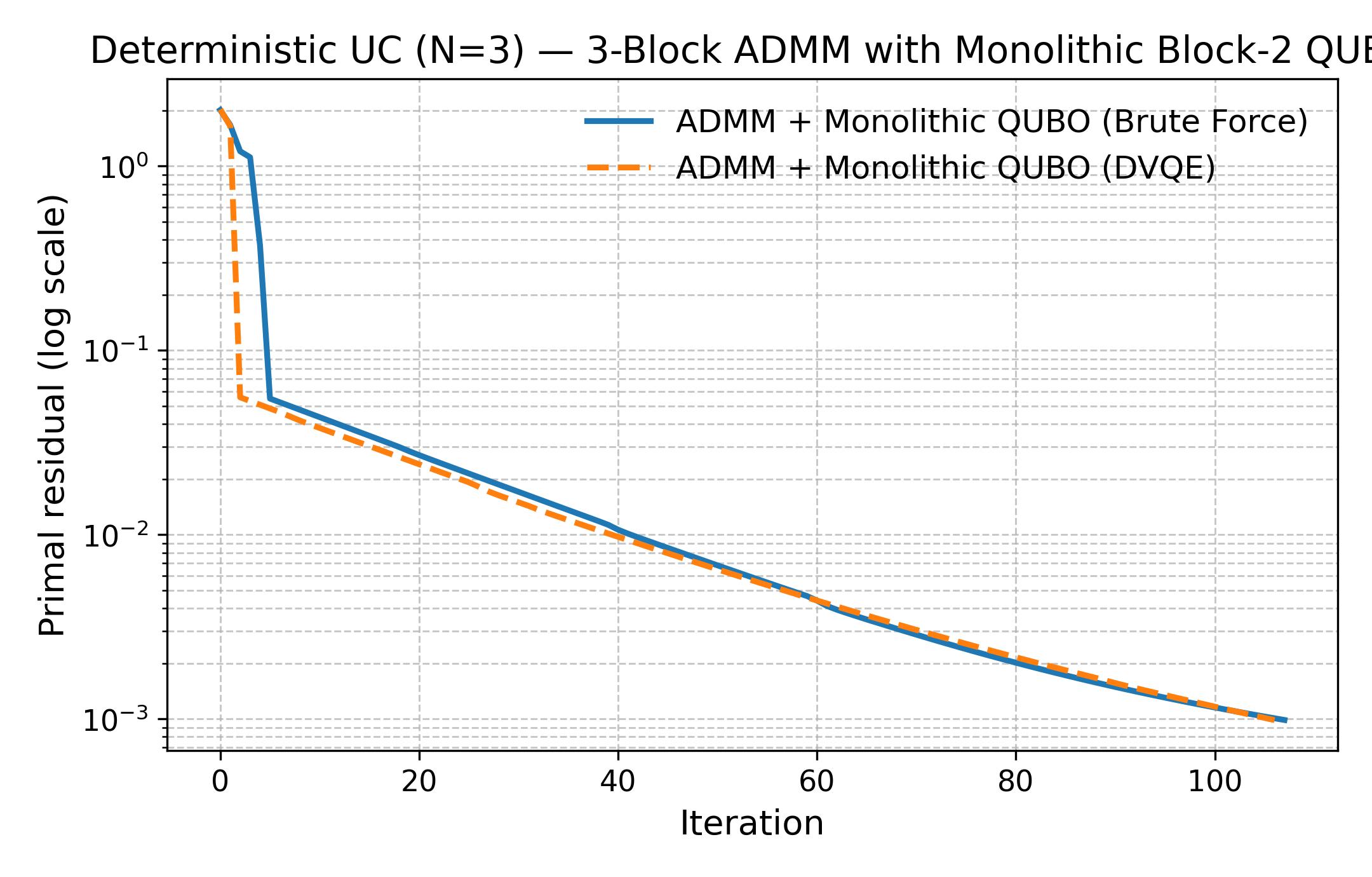}
        \caption{\scriptsize Primal residual: DVQE vs brute force}
        \label{fig:dvqe_vs_bruteforce}
    \end{subfigure}
    \hfill
    \begin{subfigure}[b]{0.48\linewidth}
        \centering
        \includegraphics[width=\linewidth]{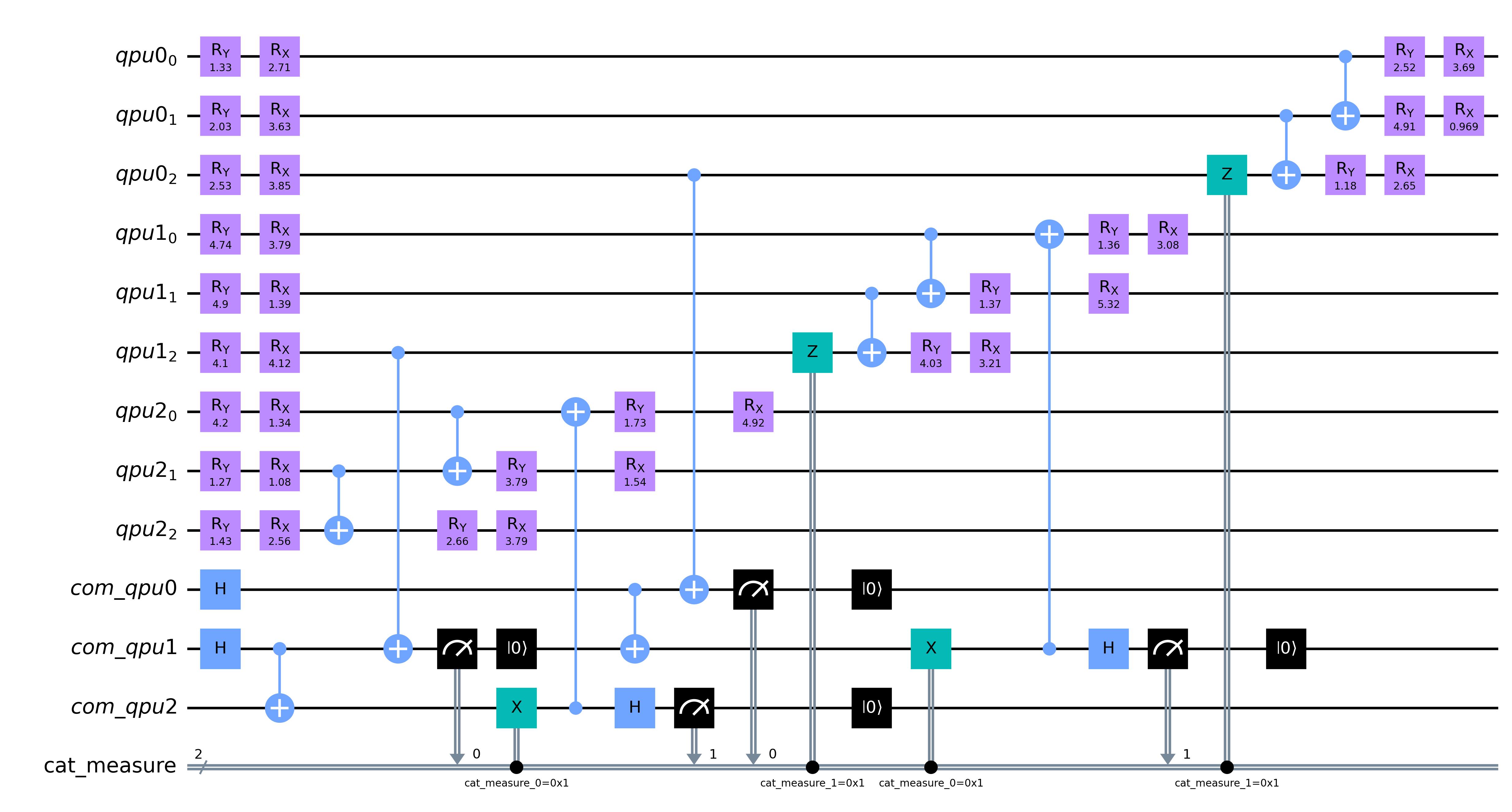}
        \caption{\scriptsize Trained DVQE ansatz circuit}
        \label{fig:dvqe_trained_ansatz}
    \end{subfigure}
    \caption{\footnotesize Validation on three-unit UC ($T{=}6$). DVQE matches brute-force optimality while reducing simulation effort.}
    \label{fig:dvqe_small}
\end{figure}

As shown in Fig.~\ref{fig:dvqe_vs_bruteforce}, the primal residual trajectories of both brute force and DVQE converge to the same threshold, confirming that DVQE attains the globally optimal binary schedule. Importantly, DVQE requires substantially less simulation time than brute force, highlighting its potential as a scalable hybrid solver. The trained ansatz circuit from the final ADMM iteration is also depicted in Fig.~\ref{fig:dvqe_trained_ansatz}. The optimized parameters reproduce the optimal commitment bitstring with high sampling probability, verifying the effectiveness of the quantum–classical training loop.  

\subsection{Five-Unit UC}

We use a five-unit UC system over $T{=}6$ hours, testing both deterministic and stochastic formulations. In each case, Block~2 is solved by brute force and DVQE for comparison. DVQE consistently reproduces the same binary schedules and objective values as brute force, but with lower runtimes.  

For the deterministic UC, three decomposition strategies are compared. The three-QUBO decomposition (separating $y$, $u$, $v$) converges to feasible schedules but requires more ADMM iterations due to weakened logical couplings. The micro-QUBO decomposition restores local dependencies and accelerates convergence, but produces a large number of trivial QUBOs, which increases overall simulation burden. The batched-QUBO strategy (three batches) strikes the best balance: it improves convergence relative to the three-QUBO case and avoids the overhead of solving thousands of microproblems, yielding batch sizes compatible with near-term quantum devices.  

\begin{figure}[!t]
    \captionsetup{font={footnotesize}}
    \centering
    \begin{subfigure}[b]{0.49\linewidth}
        \centering
        \includegraphics[width=\linewidth]{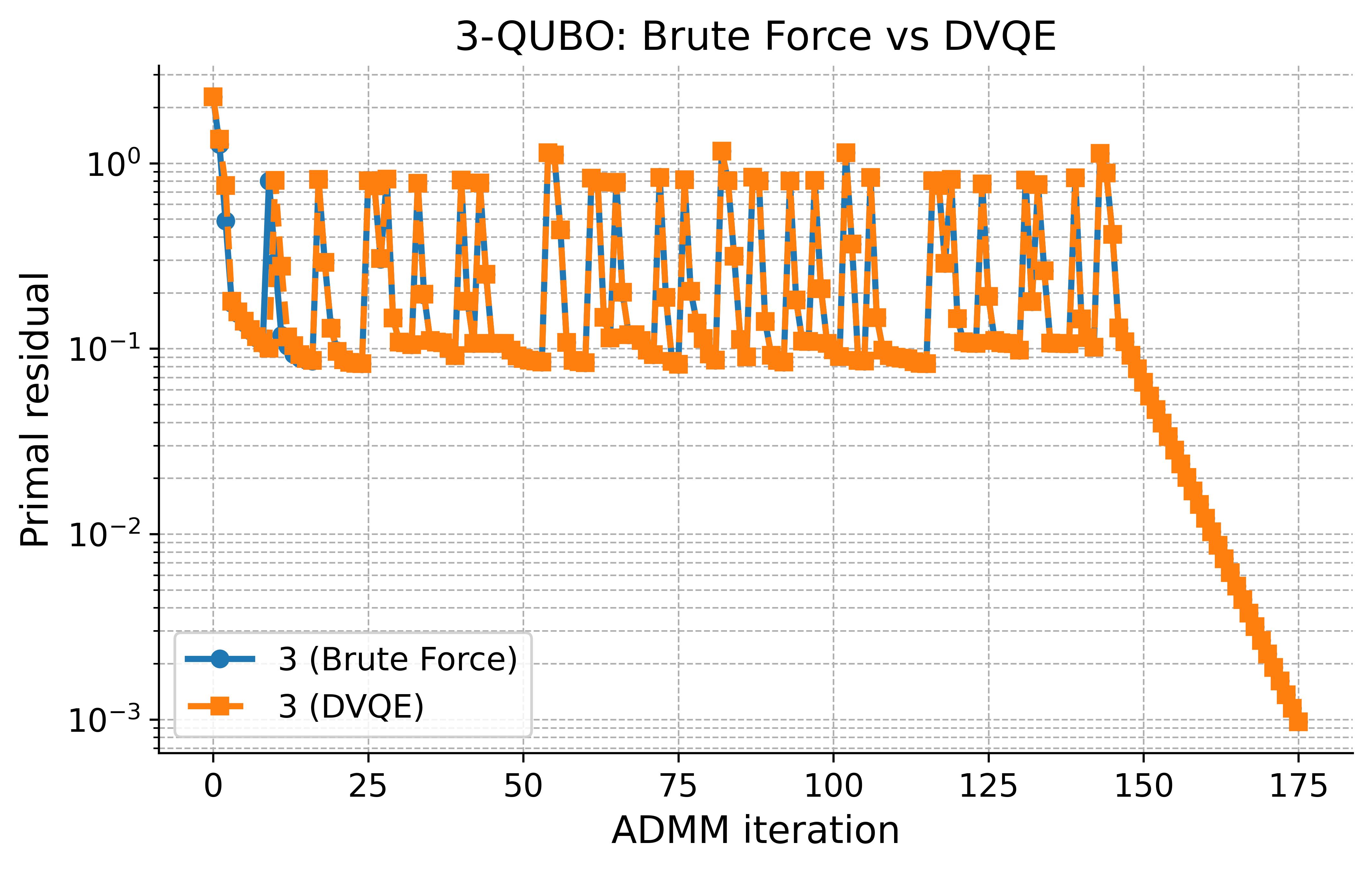}
        \caption{\scriptsize Three-QUBO decomposition}
        \label{fig:primal_det_3qubo}
    \end{subfigure}
    \hfill
    \begin{subfigure}[b]{0.49\linewidth}
        \centering
        \includegraphics[width=\linewidth]{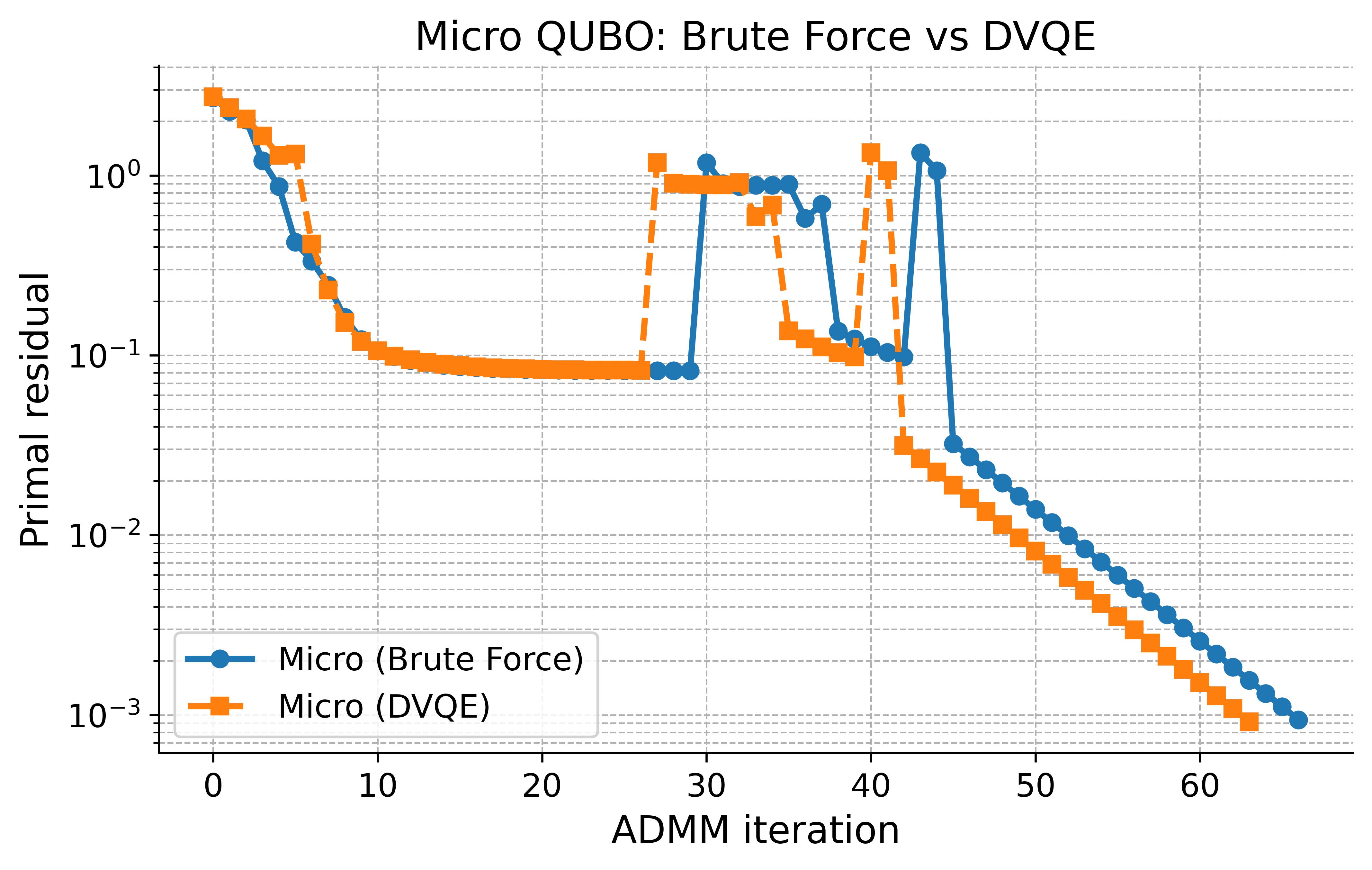}
        \caption{\scriptsize Micro-QUBO decomposition}
        \label{fig:primal_det_micro}
    \end{subfigure}

    \vspace{0.2em}

    \begin{subfigure}[b]{0.49\linewidth}
        \centering
        \includegraphics[width=\linewidth]{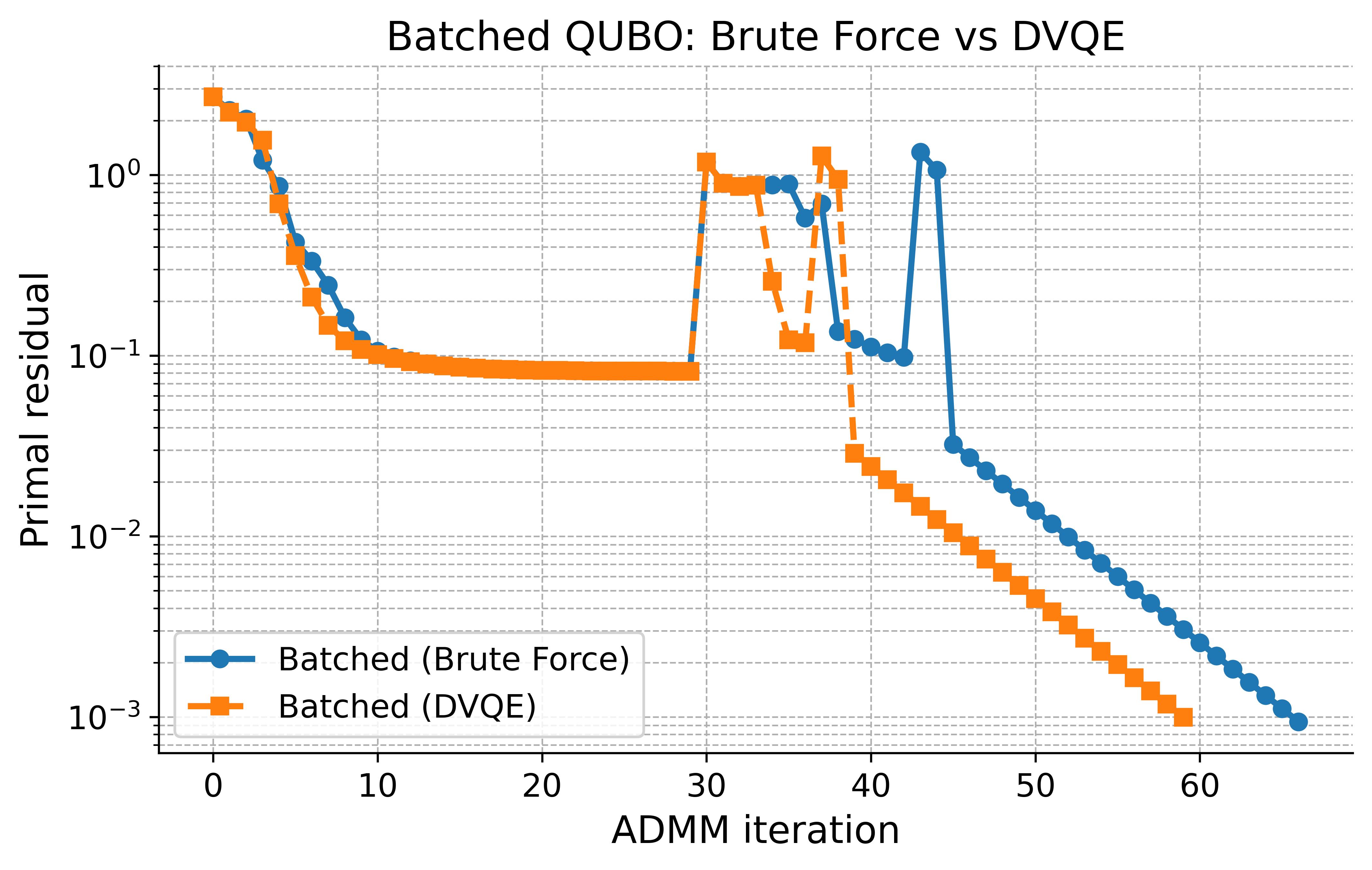}
        \caption{\scriptsize Batched-QUBO decomposition}
        \label{fig:primal_det_batched}
    \end{subfigure}
    \hfill
    \begin{subfigure}[b]{0.49\linewidth}
        \centering
        \includegraphics[width=\linewidth]{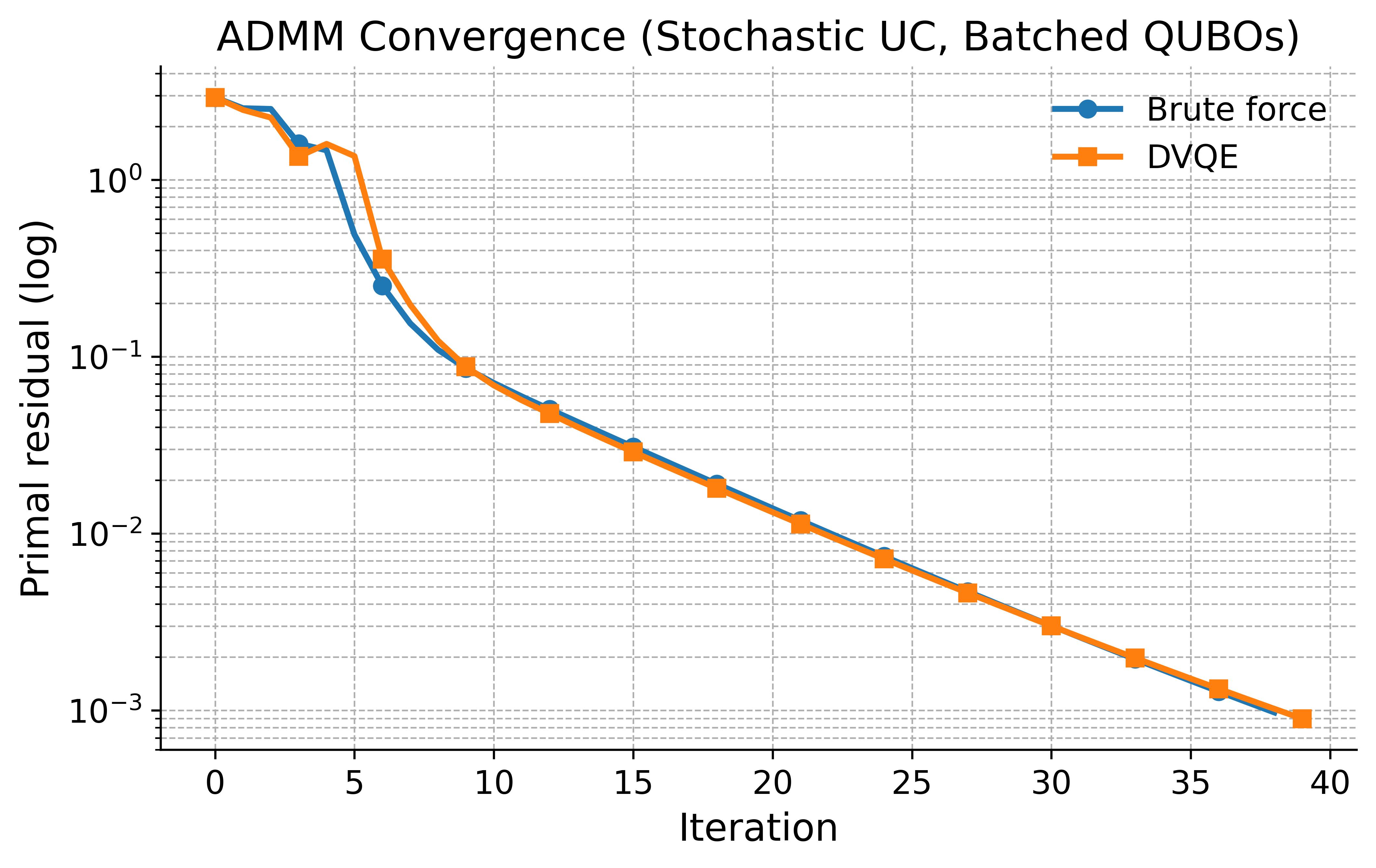}
        \caption{\scriptsize Stochastic UC with batched QUBOs}
        \label{fig:primal_stoch_batched}
    \end{subfigure}

    \caption{\footnotesize Primal residual trajectories for five-unit UC under different QUBO decompositions, solved by brute force and DVQE.}
    \label{fig:primal_residual_3badmm}
    \vspace{-9pt}
\end{figure}

Figs.~\ref{fig:primal_det_3qubo}--\ref{fig:primal_det_batched} show the primal residual trajectories for all decomposition strategies, solved by brute force and DVQE. In the micro- and batched-QUBO cases, DVQE converges faster than brute force, while in the three-QUBO case, the residual trajectories coincide exactly. Across all runs, feasibility is preserved: load balance, capacity, ramping, and reserves are satisfied; consensus is achieved; and all slack variables converge to zero. Identical ADMM hyperparameters and initialization are applied to ensure fair comparison. Table~\ref{tab:dispatch_small_clean} reports the realized deterministic dispatch.  

We next extend the study to the stochastic UC formulation under renewable uncertainty. We consider a scenario-based model with $S=4$ scenarios, $N=5$ generating units, and a planning horizon of $T=6$ periods. For this stochastic UC, we employ the batched QUBO decomposition and solve the resulting QUBOs using both brute-force enumeration and DVQE. Fig.~\ref{fig:primal_stoch_batched} shows the primal-residual trajectories for both brute force and DVQE under the batched QUBO decomposition. The residual curves coincide exactly, resulting in overlapping lines that confirm identical convergence behavior. Across both runs, we verified feasibility: power balance, capacity limits, ramping constraints, and reserve deliverability are all satisfied. Consensus is achieved, and all slack variables $(s^y,s^u,s^v)$ converge to zero. We use the same ADMM hyperparameters as in the deterministic UC case. Table~\ref{tab:stoch_dispatch} reports the realized dispatch across all units and scenarios, showing that load is met exactly at each time period in each scenario.
\begin{table}[!t]
\scriptsize  
\setlength{\tabcolsep}{4pt}  
\renewcommand{\arraystretch}{1.3}  
\captionsetup{font={footnotesize}}  
\caption{Deterministic UC ($N{=}5$, $T{=}6$): Dispatch vs.\ Demand}
\centering
\begin{tabular}{c|ccccc|c|c}
\hline
\multirow{2}{*}{\textbf{t}} 
& \multicolumn{5}{c|}{\textbf{Unit Dispatch $p_i$ (MW)}} 
& \multirow{2}{*}{$\sum_i p_{i,t}$} 
& \multirow{2}{*}{$L_t$} \\ \cline{2-6}
& $p_1$ & $p_2$ & $p_3$ & $p_4$ & $p_5$ &  &  \\ \hline
1 & 0 & 44.24 & 52.76 & 0 & 53 & 150 & 150 \\ \hline
2 & 0 & 20.00 & 70 & 0 & 80 & 170 & 170 \\ \hline
3 & 0 & 35.02 & 60 & 0 & 84.98 & 180 & 180 \\ \hline
4 & 0 & 40.02 & 35 & 0 & 84.98 & 160 & 160 \\ \hline
5 & 0 & 55.02 & 0  & 0 & 84.98 & 140 & 140 \\ \hline
6 & 0 & 45.02 & 0  & 0 & 84.98 & 130 & 130 \\ \hline
\end{tabular}
\label{tab:dispatch_small_clean}
\end{table}
\begin{table*}[!t]
\scriptsize
\setlength{\tabcolsep}{3pt}
\renewcommand{\arraystretch}{1.3}
\captionsetup{font={footnotesize}}
\caption{Stochastic UC ($N{=}5$, $T{=}6$, $S{=}4$): Dispatch vs.\ Demand Across Scenarios}
\centering
\begin{tabular}{c c}
\begin{tabular}{c|ccccc|c|c}
\hline
\multicolumn{8}{c}{\textbf{Scenario 1}} \\ \hline
\multirow{2}{*}{\textbf{t}}
& \multicolumn{5}{c|}{\textbf{Unit Dispatch $p_i$ (MW)}} 
& \multirow{2}{*}{$\sum_i p_{i,t}$} 
& \multirow{2}{*}{$L_t$} \\ \cline{2-6}
& $p_1$ & $p_2$ & $p_3$ & $p_4$ & $p_5$ & & \\ \hline
1 & 50 & 20 & 45.00 & 0 & 25 & 140 & 140 \\ \hline
2 & 50.02 & 20 & 69.99 & 0 & 20 & 160 & 160 \\ \hline
3 & 60 & 20 & 69.99 & 0 & 20.02 & 170 & 170 \\ \hline
4 & 40 & 20 & 69.99 & 0 & 20.02 & 150 & 150 \\ \hline
5 & 0  & 20 & 69.99 & 0 & 40.02 & 130 & 130 \\ \hline
6 & 0  & 20 & 69.99 & 0 & 30.02 & 120 & 120 \\ \hline
\end{tabular}
&
\begin{tabular}{c|ccccc|c|c}
\hline
\multicolumn{8}{c}{\textbf{Scenario 2}} \\ \hline
\multirow{2}{*}{\textbf{t}}
& \multicolumn{5}{c|}{\textbf{Unit Dispatch $p_i$ (MW)}} 
& \multirow{2}{*}{$\sum_i p_{i,t}$} 
& \multirow{2}{*}{$L_t$} \\ \cline{2-6}
& $p_1$ & $p_2$ & $p_3$ & $p_4$ & $p_5$ & & \\ \hline
1 & 50 & 20 & 45 & 0 & 35 & 150 & 150 \\ \hline
2 & 60 & 20 & 69.99 & 0 & 20.02 & 170 & 170 \\ \hline
3 & 60 & 20 & 69.99 & 0 & 30.02 & 180 & 180 \\ \hline
4 & 40 & 20 & 69.99 & 0 & 30.02 & 160 & 160 \\ \hline
5 & 0  & 20 & 69.99 & 0 & 50.02 & 140 & 140 \\ \hline
6 & 0  & 20 & 69.99 & 0 & 40.02 & 130 & 130 \\ \hline
\end{tabular}
\\[1ex]
\begin{tabular}{c|ccccc|c|c}
\hline
\multicolumn{8}{c}{\textbf{Scenario 3}} \\ \hline
\multirow{2}{*}{\textbf{t}}
& \multicolumn{5}{c|}{\textbf{Unit Dispatch $p_i$ (MW)}} 
& \multirow{2}{*}{$\sum_i p_{i,t}$} 
& \multirow{2}{*}{$L_t$} \\ \cline{2-6}
& $p_1$ & $p_2$ & $p_3$ & $p_4$ & $p_5$ & & \\ \hline
1 & 50 & 20 & 45 & 0 & 45 & 160 & 160 \\ \hline
2 & 60 & 20 & 69.99 & 0 & 30.02 & 180 & 180 \\ \hline
3 & 60 & 20 & 69.99 & 0 & 40.02 & 190 & 190 \\ \hline
4 & 40 & 20 & 69.99 & 0 & 40.02 & 170 & 170 \\ \hline
5 & 0  & 20 & 69.99 & 0 & 60.02 & 150 & 150 \\ \hline
6 & 0  & 20 & 69.99 & 0 & 50.02 & 140 & 140 \\ \hline
\end{tabular}
&
\begin{tabular}{c|ccccc|c|c}
\hline
\multicolumn{8}{c}{\textbf{Scenario 4}} \\ \hline
\multirow{2}{*}{\textbf{t}}
& \multicolumn{5}{c|}{\textbf{Unit Dispatch $p_i$ (MW)}} 
& \multirow{2}{*}{$\sum_i p_{i,t}$} 
& \multirow{2}{*}{$L_t$} \\ \cline{2-6}
& $p_1$ & $p_2$ & $p_3$ & $p_4$ & $p_5$ & & \\ \hline
1 & 50 & 25 & 45 & 0 & 50 & 170 & 170 \\ \hline
2 & 60 & 20 & 69.99 & 0 & 40.02 & 190 & 190 \\ \hline
3 & 60 & 20 & 69.99 & 0 & 50.02 & 200 & 200 \\ \hline
4 & 40 & 20 & 69.99 & 0 & 50.02 & 180 & 180 \\ \hline
5 & 0  & 20 & 69.99 & 0 & 70.02 & 160 & 160 \\ \hline
6 & 0  & 20 & 69.99 & 0 & 60.02 & 150 & 150 \\ \hline
\end{tabular}
\end{tabular}
\label{tab:stoch_dispatch}
\end{table*}

\subsection{Medium- and Large-Scale UC}\label{sec:medium_large}

We extend the study to larger deterministic UC problems. Two test cases are considered: a medium-scale system with $N{=}20$ generating units and a large-scale system with $N{=}50$ units, both scheduled over a horizon of $T{=}24$ hours. For both cases, Block~2 is solved using the batched QUBO decomposition. Each batch is executed by DVQE.  

\begin{figure}[!t]
    \captionsetup{font={footnotesize}}
    \centering
    \begin{subfigure}[b]{0.48\linewidth}
        \centering
        \includegraphics[width=\linewidth]{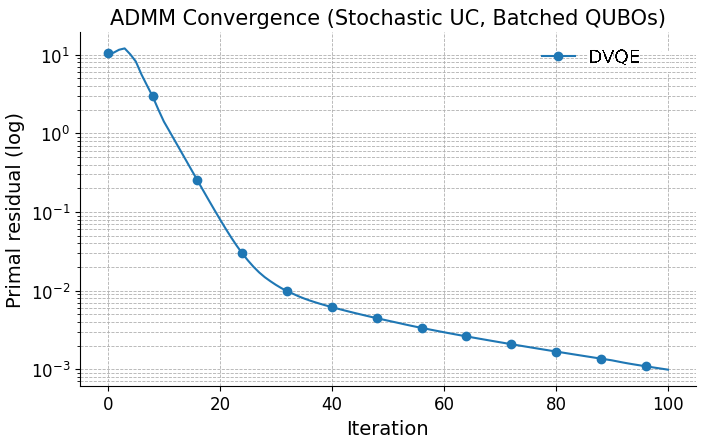}
        \caption{\scriptsize 20-unit UC ($T{=}24$): DVQE vs brute force}
        \label{fig:dvqe_medium}
    \end{subfigure}
    \hfill
    \begin{subfigure}[b]{0.48\linewidth}
        \centering
        \includegraphics[width=\linewidth]{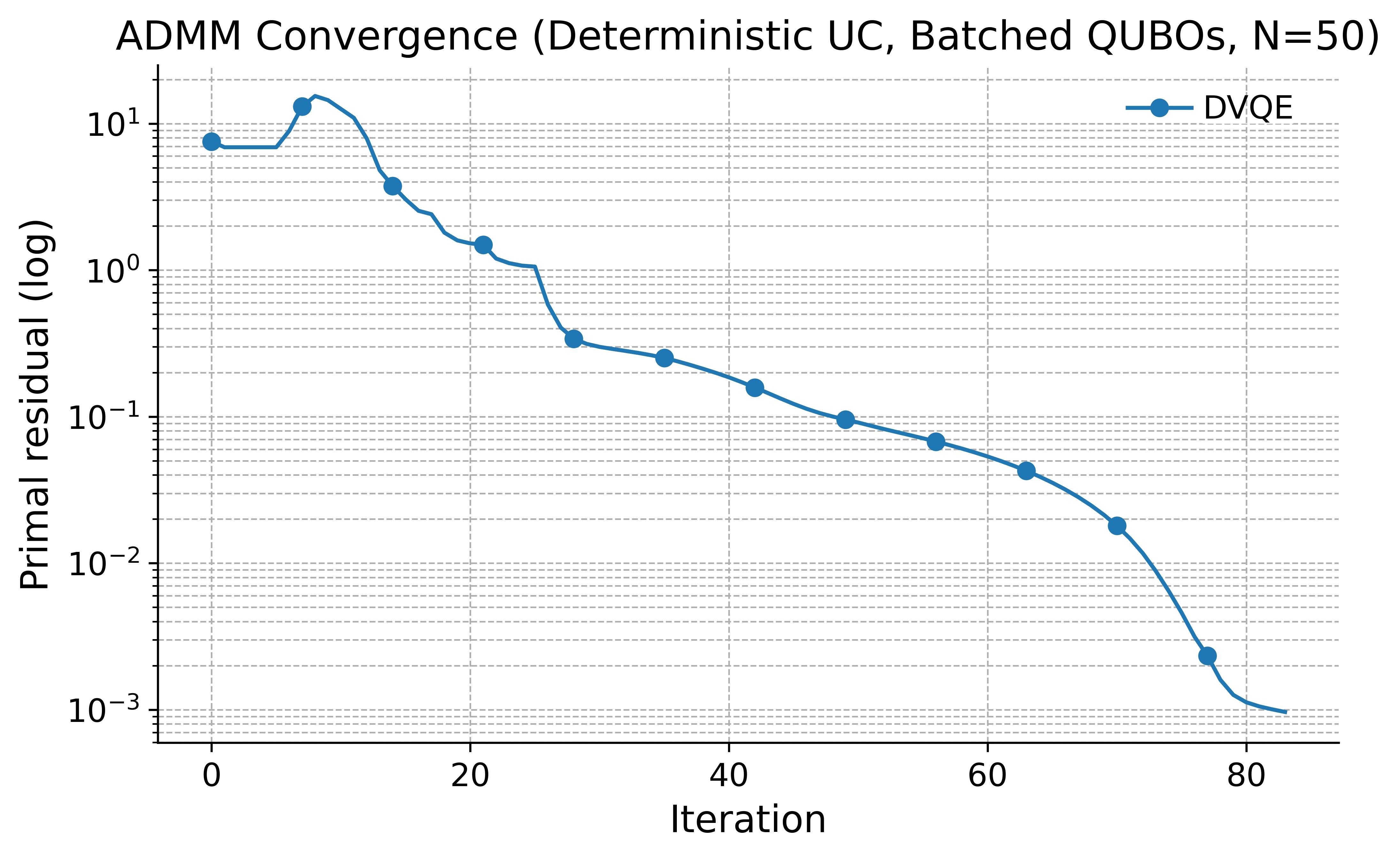}
        \caption{\scriptsize 50-unit UC ($T{=}24$): DVQE vs brute force}
        \label{fig:dvqe_large}
    \end{subfigure}
    \caption{\footnotesize Primal residual convergence for medium- and large-scale deterministic UC. DVQE and brute force coincide in residual accuracy, while DVQE runs faster in simulation time.}
    \label{fig:dvqe_medium_large}
\end{figure}

Fig.~\ref{fig:dvqe_medium_large} shows the primal-residual trajectories for the two systems. Tables~\ref{tab:dispatch_medium} and \ref{tab:dispatch_large} report partial dispatch schedules for the medium- and large-scale experiments. The complete dispatch data for all units and time periods are available in the open-source repository \href{https://github.com/LSU-RAISE-LAB/3B-ADMM-UC-DVQE}{GitHub}.  
\begin{table}[!t]
\scriptsize
\setlength{\tabcolsep}{2pt}
\renewcommand{\arraystretch}{1.2}
\captionsetup{font={footnotesize}}
\caption{Partial Dispatch for Medium-Scale UC ($N{=}20$, $T{=}24$). Ten representative units over first six hours.}
\centering
\begin{tabular}{c|cccccccccc|c|c}
\hline
\multirow{2}{*}{\textbf{t}} & \multicolumn{10}{c|}{\textbf{Unit Dispatch $p_i$ (MW)}} & \multirow{2}{*}{$\sum_i p_{i,t}$} & \multirow{2}{*}{$L_t$} \\ \cline{2-11}
& $p_1$ & $p_3$ & $p_4$ & $p_6$ & $p_7$ & $p_{10}$ & $p_{12}$ & $p_{15}$ & $p_{18}$ & $p_{20}$ & & \\ \hline
1 & 0 & 70 & 0 & 70 & 85 & 60 & 0 & 85 & 70 & 80 & 900 & 900 \\ \hline
2 & 0 & 100 & 0 & 40 & 40 & 30 & 0 & 40 & 100 & 110 & 950 & 950 \\ \hline
3 & 33.3 & 70 & 0 & 30 & 35 & 30 & 50 & 35 & 130 & 70 & 1000 & 1000 \\ \hline
4 & 76.7 & 40 & 20 & 30 & 35 & 30 & 90 & 35 & 130 & 40 & 1050 & 1050 \\ \hline
5 & 100 & 30 & 20 & 30 & 35 & 30 & 100 & 35 & 140 & 40 & 1100 & 1100 \\ \hline
6 & 100 & 43.1 & 20 & 30 & 35 & 30 & 100 & 35 & 140 & 40 & 1150 & 1150 \\ \hline
\end{tabular}
\label{tab:dispatch_medium}
\end{table}
\begin{table}[!t]
\scriptsize
\setlength{\tabcolsep}{2.5pt}
\renewcommand{\arraystretch}{1.2}
\captionsetup{font={footnotesize}}
\caption{Partial Dispatch for UC ($N{=}50$, $T{=}24$). Ten representative units over first six hours.}
\centering
\begin{tabular}{c|cccccccccc|c|c}
\hline
\multirow{2}{*}{\textbf{t}} & \multicolumn{10}{c|}{\textbf{Unit Dispatch $p_i$ (MW)}} & \multirow{2}{*}{$\sum_i p_{i,t}$} & \multirow{2}{*}{$L_t$} \\ \cline{2-11}
& $p_2$ & $p_5$ & $p_{9}$ & $p_{11}$ & $p_{14}$ & $p_{20}$ & $p_{25}$ & $p_{33}$ & $p_{42}$ & $p_{50}$ & & \\ \hline
1 & 0 & 66.7 & 17.9 & 0 & 0 & 51.5 & 66.8 & 0 & 0 & 52.8 & 900 & 900 \\ \hline
2 & 0 & 66.7 & 41 & 0 & 0 & 44.1 & 66.8 & 0 & 0 & 52.8 & 950 & 950 \\ \hline
3 & 24.8 & 42.9 & 17.9 & 0 & 27.9 & 19.6 & 66.8 & 0 & 13.8 & 68.4 & 1000 & 1000 \\ \hline
4 & 49.7 & 19.1 & 19 & 0 & 51.4 & 19.6 & 66.8 & 15.8 & 16.9 & 53.4 & 1050 & 1050 \\ \hline
5 & 52 & 19.1 & 19 & 0 & 46 & 19.6 & 66.8 & 17.7 & 16.9 & 47.9 & 1100 & 1100 \\ \hline
6 & 39.1 & 19.1 & 19 & 0 & 34.6 & 19.6 & 66.8 & 17.7 & 16.9 & 36 & 1150 & 1150 \\ \hline
\end{tabular}
\label{tab:dispatch_large}
\end{table}

\section{Conclusion}\label{sec:conclusion}
This paper proposed D²-UC, a quantum-ready reformulation of the UC problem using a three-block ADMM decomposition that isolates all binary decisions in a QUBO subproblem. To make this block compatible with near-term hardware, we advanced a progression of strategies: from a monolithic QUBO, to three type-specific QUBOs, to micro-QUBOs per unit–time pair, and finally to batched micro-QUBOs that reduce a large number of subproblems to a fixed number of solver-ready instances. A distributed quantum solver was presented to solve the decomposed QUBOs in a distributed quantum setting. An accept-if-better safeguard was also integrated with the DVQE to stabilize hybrid updates. Case studies on deterministic and stochastic UC demonstrated that the framework delivers feasible schedules, faster ADMM convergence, and QUBO sizes well suited to current hybrid and distributed quantum solvers.

\bibliographystyle{IEEEtran}
\bibliography{example}

@article{hasanzadeh2025distributed,
  title={Distributed Implementation of Variational Quantum Eigensolver to Solve {QUBO} Problems},
  author={Hasanzadeh, Milad and Kargarian, Amin},
  journal={arXiv preprint arXiv:2508.17471},
  year={2025}
}

@article{carrion2006computationally,
  title={A computationally efficient mixed-integer linear formulation for the thermal unit commitment problem},
  author={Carri{\'o}n, Miguel and Arroyo, Jos{\'e} M},
  journal={IEEE Transactions on power systems},
  volume={21},
  number={3},
  pages={1371--1378},
  year={2006},
  publisher={IEEE}
}

@article{padhy2004unit,
  title={Unit commitment-a bibliographical survey},
  author={Padhy, Narayana Prasad},
  journal={IEEE Transactions on power systems},
  volume={19},
  number={2},
  pages={1196--1205},
  year={2004},
  publisher={IEEE}
}

@article{asensio2015stochastic,
  title={Stochastic unit commitment in isolated systems with renewable penetration under CVaR assessment},
  author={Asensio, Miguel and Contreras, Javier},
  journal={IEEE Transactions on Smart Grid},
  volume={7},
  number={3},
  pages={1356--1367},
  year={2015},
  publisher={IEEE}
}

@article{mahroo2023learning,
  title={Learning infused quantum-classical distributed optimization technique for power generation scheduling},
  author={Mahroo, Reza and Kargarian, Amin},
  journal={IEEE Transactions on Quantum Engineering},
  volume={4},
  pages={1--14},
  year={2023},
  publisher={IEEE}
}

@article{nikmehr2022quantum,
  title={Quantum distributed unit commitment: An application in microgrids},
  author={Nikmehr, Nima and Zhang, Peng and Bragin, Mikhail A},
  journal={IEEE transactions on power systems},
  volume={37},
  number={5},
  pages={3592--3603},
  year={2022},
  publisher={IEEE}
}

@article{farhi2014quantum,
  title={A quantum approximate optimization algorithm},
  author={Farhi, Edward and Goldstone, Jeffrey and Gutmann, Sam},
  journal={arXiv preprint arXiv:1411.4028},
  year={2014}
}

@article{preskill2018quantum,
  title={Quantum computing in the {NISQ} era and beyond},
  author={Preskill, John},
  journal={Quantum},
  volume={2},
  pages={79},
  year={2018},
  publisher={Verein zur F{\"o}rderung des Open Access Publizierens in den Quantenwissenschaften}
}

@article{peruzzo2014variational,
  title={A variational eigenvalue solver on a photonic quantum processor},
  author={Peruzzo, Alberto and McClean, Jarrod and Shadbolt, Peter and Yung, Man-Hong and Zhou, Xiao-Qi and Love, Peter J and Aspuru-Guzik, Al{\'a}n and O’brien, Jeremy L},
  journal={Nature communications},
  volume={5},
  number={1},
  pages={4213},
  year={2014},
  publisher={Nature Publishing Group UK London}
}

@book{conejo2006decomposition,
  title={Decomposition techniques in mathematical programming: engineering and science applications},
  author={Conejo, Antonio J and Castillo, Enrique and Minguez, Roberto and Garcia-Bertrand, Raquel},
  year={2006},
  publisher={Springer}
}

@article{han2025quantum,
  title={Quantum Computing for Stochastic Economic Dispatch in Renewables-Rich Power Systems},
  author={Han, Xutao and Li, Zhiyi and Gu, Wei and Shahidehpour, Mohammad},
  journal={IEEE Transactions on Smart Grid},
  year={2025},
  publisher={IEEE}
}

@article{glover2018tutorial,
  title={A tutorial on formulating and using {QUBO} models},
  author={Glover, Fred and Kochenberger, Gary and Du, Yu},
  journal={arXiv preprint arXiv:1811.11538},
  year={2018}
}

@article{caleffi2024distributed,
  title={Distributed quantum computing: a survey},
  author={Caleffi, Marcello and Amoretti, Michele and Ferrari, Davide and Illiano, Jessica and Manzalini, Antonio and Cacciapuoti, Angela Sara},
  journal={Computer Networks},
  volume={254},
  pages={110672},
  year={2024},
  publisher={Elsevier}
}

@article{diadamo2021distributed,
  title={Distributed quantum computing and network control for accelerated {VQE}},
  author={DiAdamo, Stephen and Ghibaudi, Marco and Cruise, James},
  journal={IEEE Transactions on Quantum Engineering},
  volume={2},
  pages={1--21},
  year={2021},
  publisher={IEEE}
}

@article{du2022distributed,
  title={A distributed learning scheme for variational quantum algorithms},
  author={Du, Yuxuan and Qian, Yang and Wu, Xingyao and Tao, Dacheng},
  journal={IEEE Transactions on Quantum Engineering},
  volume={3},
  pages={1--16},
  year={2022},
  publisher={IEEE}
}

@inproceedings{yimsiriwattana2004distributed,
  title={Distributed quantum computing: A distributed Shor algorithm},
  author={Yimsiriwattana, Anocha and Lomonaco Jr, Samuel J},
  booktitle={Quantum information and computation II},
  volume={5436},
  pages={360--372},
  year={2004},
  organization={SPIE}
}

@article{11177244,
  title={{ADMM} Enhancement Techniques for Distributed Optimal Power Flow},
  author={Hasanzadeh, Milad and Kargarian, Amin},
  journal={IEEE Transactions on Power Systems},
  year={2025},
  publisher={IEEE}
}

@article{zhang2020two,
  title={Two-stage fully distributed approach for unit commitment with consensus {ADMM}},
  author={Zhang, Chen and Yang, Linfeng and Jian, Jinbao},
  journal={Electric Power Systems Research},
  volume={181},
  pages={106180},
  year={2020},
  publisher={Elsevier}
}

@article{shi2025two,
  title={Two-Stage Linearization of Frequency Nadir Constraint for Unit Commitment},
  author={Shi, Qingxin and Zhu, Yanren and Fan, Ke and Cheng, Rui and Shen, Jialin and Liu, Wenxia},
  journal={IEEE Transactions on Power Systems},
  year={2025},
  publisher={IEEE}
}

@article{dong2025data,
  title={A Data-Driven Cost Budget Satisficing Model for Unit Commitment under Solar Power Uncertainty},
  author={Dong, Hanjiang and Wu, Lubin and Zhu, Jizhong and Li, Shenglin and Liang, Zipeng and Yang, Haosen and Chung, Chi-yung},
  journal={IEEE Transactions on Power Systems},
  year={2025},
  publisher={IEEE}
}

@article{stein2305combining,
  title={Combining the qaoa and hhl algorithm to achieve a substantial quantum speedup for the unit commitment problem (2023)},
  author={Stein, J and Jojo, J and Farea, A and Bucher, D and Altmann, P and {\c{C}}elebi, MS and Linnhoff-Popien, C},
  journal={arXiv preprint arXiv:2305.08482}
}

@article{kingma2014adam,
  title={{ADAM}: A method for stochastic optimization},
  author={Kingma, Diederik P},
  journal={arXiv preprint arXiv:1412.6980},
  year={2014}
}

@article{pareek2025limitations,
  title={Limitations of Fault-Tolerant Quantum Linear System Solvers for Quantum Power Flow},
  author={Pareek, Parikshit and Jayakumar, Abhijith and Coffrin, Carleton and Misra, Sidhant},
  journal={IEEE Transactions on Power Systems},
  year={2025},
  publisher={IEEE}
}

@article{morstyn2022annealing,
  title={Annealing-based quantum computing for combinatorial optimal power flow},
  author={Morstyn, Thomas},
  journal={IEEE Transactions on Smart Grid},
  volume={14},
  number={2},
  pages={1093--1102},
  year={2022},
  publisher={IEEE}
}

@article{javadi2025learning,
  title={Learning Constraint Surrogate Model for Two-stage Stochastic Unit Commitment},
  author={Javadi, Amir Bahador and Kargarian, Amin and Naraghi-Pour, Mort},
  journal={arXiv preprint arXiv:2509.10246},
  year={2025}
}

@article{gambella2020multiblock,
  title={Multiblock ADMM heuristics for mixed-binary optimization on classical and quantum computers},
  author={Gambella, Claudio and Simonetto, Andrea},
  journal={IEEE Transactions on Quantum Engineering},
  volume={1},
  pages={1--22},
  year={2020},
  publisher={IEEE}
}

\end{document}